\documentclass[12pt]{article}
\usepackage{graphics,graphicx}
\usepackage{amssymb,epsfig,amsmath,euscript,array}
\usepackage{cite}
\usepackage{axodraw}
\usepackage{pstricks}
\usepackage{color}
\usepackage{units}
\usepackage[unicode=true, pdfusetitle,
 bookmarks=true,bookmarksnumbered=false,bookmarksopen=false,
 breaklinks=true,pdfborder={0 0 0},backref=false,colorlinks=false]
 {hyperref}

\makeatletter
\@addtoreset{equation}{section}
\makeatother

\newcounter{multieqs}

\newcommand{\be}{\begin{equation}}
\newcommand{\ee}{\end{equation}}

\newcommand{\kslash}{k \!\!\! / }

\newcommand{\lslash}{l \!\! / }
\newcommand{\Pslash}{P \!\!\!\! / }

\newcommand{\islash}{i \!\!\! / }
\newcommand{\jslash}{j \!\!\! / }
\newcommand{\aslash}{a \!\!\! / }
\newcommand{\bslash}{{b \hspace{-6pt} \slash} }

\newcommand{\onslash}{1 \!\!\! / }
\newcommand{\twslash}{2 \!\!\!/ }
\newcommand{\thslash}{3 \!\!\!/ }
\newcommand{\foslash}{4 \!\!\! / }
\newcommand{\fislash}{5 \!\!\! / }

\newcommand{\mslash}{m \!\!\! / }

\def\bd{\begin{document}}
\def\ed{\end{document}}
\def\nn{\nonumber}
\def\bea{\begin{eqnarray}}
\def\eea{\end{eqnarray}}

\def\ab{(ijab)}
\def\ba{(ijba)}
\def\ijab{{\tr}_{+}(\islash\, \jslash\, \aslash \, \bslash)}
\def\ijba{{\tr}_{+}(\islash\, \jslash\, \bslash \, \aslash)}
\def\ijaP{{\tr}_{+}(\islash\, \jslash\, \aslash \, \Pslash)}
\def\ijPLa{{\tr}_{+}(\islash\, \jslash\, \Pslash_L \, \aslash)}
\def\ijaPL{{\tr}_{+}(\islash\, \jslash\, \aslash \, \Pslash_L)}
\def\ijPLza{{\tr}_{+}(\islash\, \jslash\, \Pslash_{L;z} \, \aslash)}
\def\ijaPLz{{\tr}_{+}(\islash\, \jslash\, \aslash \, \Pslash_{L;z})}
\def\ijPa{{\tr}_{+}(\islash\, \jslash\, \Pslash \, \aslash)}
\def\iaPb{{\tr}_{+}(\islash\, \aslash\, \Pslash \, \bslash)}
\def\ibPa{{\tr}_{+}(\islash\, \bslash\, \Pslash \, \aslash)}
\def\ijPmu{{\tr}_{+}(\islash\, \jslash\, \Pslash \, \mu)}
\def\ibmuP{{\tr}_{+}(\islash\, \bslash\, \mu \, \Pslash)}
\def\ibmua{{\tr}_{+}(\islash\, \bslash\, \mu \, \aslash)}
\def\iamub{{\tr}_{+}(\islash\, \aslash\, \mu \, \bslash)}
\def\jaPb{{\tr}_{+}(\jslash\, \aslash\, \Pslash \, \bslash)}
\def\ijmuP{{\tr}_{+}(\islash\, \jslash\, \mu \, \Pslash)}
\def\ijmum{{\tr}_{+}(\islash\, \jslash\, \mu \, \mslash)}
\def\ijmmu{{\tr}_{+}(\islash\, \jslash\, \mslash \, \mu)}
\def\ijmP{{\tr}_{+}(\islash\, \jslash\, \mslash \, \Pslash)}
\def\iabP{{\tr}_{+}(\islash\, \aslash\, \bslash \, \Pslash)}
\def\ijbP{{\tr}_{+}(\islash\, \jslash\, \bslash \, \Pslash)}
\def\jbPa{{\tr}_{+}(\jslash\, \bslash\, \Pslash \, \aslash)}
\def\ijPb{{\tr}_{+}(\islash\, \jslash\, \Pslash \, \bslash)}
\def\jbmua{{\tr}_{+}(\jslash\, \bslash\, \mu \, \aslash)}

\def\loablt{ {\tr}_{+}(\lslash_1\, \aslash \, \bslash\, \lslash_2)}

\def\ijlolt{{\tr}_{+}(\islash\, \jslash\, \lslash_1 \, \lslash_2)}
\def\ijltlo{{\tr}_{+}(\islash\, \jslash\, \lslash_2 \, \lslash_1)}
\def\ibloa{{\tr}_{+}(\islash\, \bslash\, \lslash_1 \, \aslash)}
\def\jaltb{{\tr}_{+}(\jslash\, \aslash\, \lslash_2 \, \bslash)}
\def\ialtb{{\tr}_{+}(\islash\, \aslash\, \lslash_2 \, \bslash)}
\def\bltloa{{\tr}_{+}(\bslash\, \lslash_2\, \lslash_1 \, \aslash)}
\def\jbloa{{\tr}_{+}(\jslash\, \bslash\, \lslash_1 \, \aslash)}
\def\ibPb{{\tr}_{+}(\islash\, \bslash\, \Pslash \, \bslash)}
\def\ijltb{{\tr}_{+}(\islash\, \jslash\, \lslash_2 \, \bslash)}

\def\ijloa{{\tr}_{+}(\islash\, \jslash\,  \lslash_1 \, \aslash)}
\def\ijblt{{\tr}_{+}(\islash\, \jslash\,  \bslash \, \lslash_2)}

\def\jakb{{\tr}_{+}(\jslash\, \aslash\, \kslash \, \bslash)}
\def\iakb{{\tr}_{+}(\islash\, \aslash\, \kslash \, \bslash)}

\def\tofo{{\tr}_{+}(\onslash\, \thslash\, \twslash \, \foslash)}
\def\foto{{\tr}_{+}(\onslash\, \thslash\, \foslash \, \twslash)}
\def\tofi{{\tr}_{+}(\onslash\, \thslash\, \twslash \, \fislash)}
\def\fito{{\tr}_{+}(\onslash\, \thslash\, \fislash \, \twslash)}

\def\lrangle#1#2{\langle #1\,#2\rangle}

\def\Li{{$\rm Li}_2$}
\def\eps{\epsilon}
\def\epsuv{{\epsilon_{\rm \mbox{\tiny UV}}}}
\let\bm=\bibitem

\def\npb#1#2#3{Nucl. Phys. {\bf{B#1}} #3 (#2)}
\def\plb#1#2#3{Phys. Lett. {\bf{#1B}} #3 (#2)}
\def\prl#1#2#3{Phys. Rev. Lett. {\bf{#1}} #3 (#2)}
\def\prd#1#2#3{Phys. Rev. {D \bf{#1}} #3 (#2)}
\def\cmp#1#2#3{Comm. Math. Phys. {\bf{#1}} #3 (#2)}
\def\cqg#1#2#3{Class. Quantum Grav. {\bf{#1}} #3 (#2)}
\def\nppsa#1#2#3{Nucl. Phys. B (Proc. Suppl.) {\bf{#1A}}#3 (#2)}
\def\ap#1#2#3{Ann. of Phys. {\bf{#1}} #3 (#2)}
\def\ijmp#1#2#3{Int. J. Mod. Phys. {\bf{A#1}} #3 (#2)}
\def\rmp#1#2#3{Rev. Mod. Phys. {\bf{#1}} #3 (#2)}
\def\mpla#1#2#3{Mod. Phys. Lett. {\bf A#1} #3 (#2)}
\def\jhep#1#2#3{J. High Energy Phys. {\bf #1} #3 (#2)}
\def\atmp#1#2#3{Adv. Theor. Math. Phys. {\bf #1} #3 (#2)}
\newcommand{\EQ}[1]{\begin{equation} #1 \end{equation}}
\newcommand{\AL}[1]{\begin{subequations}\begin{align} #1 \end{align}\end{subequations}}
\newcommand{\SP}[1]{\begin{equation}\begin{split} #1 \end{split}\end{equation}}
\newcommand{\ALAT}[2]{\begin{subequations}\begin{alignat}{#1} #2 \end{alignat}
                        \end{subequations}}
\def\beqa{\begin{eqnarray}}
\def\eeqa{\end{eqnarray}}
\def\beq{\begin{equation}}
\def\eeq{\end{equation}}
\def\sst{\scriptscriptstyle}
\def\thetabar{\bar\theta}
\def\Tr{{\rm Tr}}
\def\one{\mbox{1 \kern-.59em {\rm l}}}
 \def\Nh{\hat{N}}

\newcommand{\half}{{\textstyle {1 \over 2}}}

\def\cA{{\cal A}} \def\cB{{\cal B}} \def\cC{{\cal C}}
\def\cD{{\cal D}} \def\cE{{\cal E}} \def\cF{{\cal F}}
\def\cG{{\cal G}} \def\cH{{\cal H}} \def\cI{{\cal I}}
\def\cJ{{\cal J}} \def\cK{{\cal K}} \def\cL{{\cal L}}
\def\cM{{\cal M}} \def\cN{{\cal N}} \def\cO{{\cal O}}
\def\cP{{\cal P}} \def\cQ{{\cal Q}} \def\cR{{\cal R}}
\def\cS{{\cal S}} \def\cT{{\cal T}} \def\cU{{\cal U}}
\def\cV{{\cal V}} \def\cW{{\cal W}} \def\cX{{\cal X}}
\def\cY{{\cal Y}} \def\cZ{{\cal Z}}
\def\ua{\underline{\alpha}}
\def\ub{\underline{\phantom{\alpha}}\!\!\!\beta}
\def\uc{\underline{\phantom{\alpha}}\!\!\!\gamma}
\def\um{\underline{\mu}}
\def\ud{\underline\delta}
\def\ue{\underline\epsilon}
\def\una{\underline a}\def\unA{\underline A}
\def\unb{\underline b}\def\unB{\underline B}
\def\unc{\underline c}\def\unC{\underline C}
\def\und{\underline d}\def\unD{\underline D}
\def\une{\underline e}\def\unE{\underline E}
\def\unf{\underline{\phantom{e}}\!\!\!\! f}\def\unF{\underline F}
\def\unm{\underline m}\def\unM{\underline M}
\def\unn{\underline n}\def\unN{\underline N}
\def\unp{\underline{\phantom{a}}\!\!\! p}\def\unP{\underline P}
\def\unq{\underline{\phantom{a}}\!\!\! q}
\def\unQ{\underline{\phantom{A}}\!\!\!\! Q}
\def\unH{\underline{H}}
\def\As {{A \hspace{-6.4pt} \slash}\;}
\def\bs {{b \hspace{-6.4pt} \slash}\;}
\def\Ds {{D \hspace{-6.4pt} \slash}\;}
\def\ds {{\del \hspace{-6.4pt} \slash}\;}
\def\ss {{\s \hspace{-6.4pt} \slash}\;}
\def\ks {{ k \hspace{-6.4pt} \slash}\;}
\def\ps {{p \hspace{-6.4pt} \slash}\;}
\def\pas {{{p_1} \hspace{-6.4pt} \slash}\;}
\def\pbs {{{p_2} \hspace{-6.4pt} \slash}\;}
\def\Ps {{P \hspace{-6.4pt} \slash}\;}
\def\Qs {{Q \hspace{-6.4pt} \slash}\;}
\def\Fh{\hat{F}}
\def\Vh{\hat{V}}
\def\Xh{\hat{X}}
\def\ah{\hat{a}}
\def\xh{\hat{x}}
\def\yh{\hat{y}}
\def\ph{\hat{p}}
\def\xih{\hat{\xi}}
\def\psit{\tilde{\psi}}
\def\Psit{\tilde{\Psi}}
\def\tht{\tilde{\th}}
\def\lt{\tilde{\lambda}}
\def\hl{\hat{\lambda}}
\def\hlt{\hat{\tilde{\lambda}}}
\def\llt{\tilde{l}}
\def\At{\tilde{A}}
\def\Qt{\tilde{Q}}
\def\Rt{\tilde{R}}
\def\Nt{\tilde{N}}

\def\at{\tilde{a}}
\def\st{\tilde{s}}
\def\ft{\tilde{f}}
\def\pt{\tilde{p}}
\def\qt{\tilde{q}}
\def\vt{\tilde{v}}
\def\nt{\tilde{n}}
\def\delb{\bar{\partial}}
\def\bz{\bar{z}}
\def\bD{\bar{D}}
\def\bB{\bar{B}}
\def\bk{{\bf k}}
\def\bl{{\bf l}}
\def\bp{{\bf p}}
\def\bq{{\bf q}}
\def\br{{\bf r}}
\def\bx{{\bf x}}
\def\by{{\bf y}}
\def\bR{{\bf R}}
\def\bV{{\bf V}}
\def\d{\delta}\def\D{\Delta}\def\ddt{\dot\delta}
\def\pa{\partial} \def\del{\partial}
\def\xx{\times}
\def\uno{\mbox{1 \kern-.59em {\rm l}}}
\def\trp{^{\top}}
\def\inv{^{-1}}
\def\dag{{^{\dagger}}}
\def\pr{^{\prime}}
\def\rar{\rightarrow}
\def\lar{\leftarrow}
\def\lrar{\leftrightarrow}
\newcommand{\0}{\,\!}      %this is just NOTHING!
\def\one{1\!\!1\,\,}
\def\im{\imath}
\def\jm{\jmath}
\newcommand{\tr}{\mbox{tr}}
\newcommand{\slsh}[1]{/ \!\!\!\! #1}
\def\vac{|0\rangle}
\def\hlf{\frac{1}{2}}
\def\ove#1{\frac{1}{#1}}
\def\Box{\square}
\def\ZZ{\mathbb{Z}}
\def\CC#1{({\bf #1})}
\def\bcomment#1{}
\def\bfhat#1{{\bf \hat{#1}}}
\newcommand{\ex}[1]{{\rm e}^{#1}} \def\ii{{\rm i}}
\def\rr{{\rm r}} \def\rs{{\rm s}}\def\rv{{\rm v}}
\def\ri{{\rm i}}\def\rj{{\rm j}}
\newcommand{\lrbrk}[1]{\left(#1\right)}
\newcommand{\sfrac}[2]{{\textstyle\frac{#1}{#2}}}

\def\Li{{\rm Li}_2}

\newcommand{\la}{\langle}
\newcommand{\ra}{\rangle}
\newcommand{\da}{\dot{a}}
\newcommand{\db}{\dot{b}}
\newcommand{\dc}{\dot{c}}
\newcommand{\dd}{\dot{d}}
\newcommand{\de}{\dot{e}}
\newcommand{\df}{\dot{f}}
\newcommand{\dalpha}{\dot{\alpha}}
\newcommand{\dbeta}{\dot{\beta}}
\newcommand{\dgamma}{\dot{\gamma}}
\newcommand{\ddelta}{\dot{\delta}}

\newcommand{\tlambda}{\tilde{\lambda}}
\newcommand{\teta}{\tilde{\eta}}
\newcommand{\tsigma}{\tilde{\sigma}}
\newcommand{\tu}{\tilde{u}}
\newcommand{\tw}{\tilde{w}}
\newcommand{\tW}{\tilde{W}}
\newcommand{\tG}{\tilde{\Gamma}}
\newcommand{\tD}{\tilde{\Delta}}
\newcommand{\tpsi}{\tilde{\psi}}

\newcommand{\tx}{\tilde{x}}
\newcommand{\tq}{\tilde{q}}
\newcommand{\tQ}{\tilde{Q}}

\newcommand{\btheta}{\bar{\theta}}
\newcommand{\bpsi}{\bar{\psi}}
\newcommand{\blambda}{\bar{\lambda}}

\newcommand{\hone}{\hat{1}}
\newcommand{\htwo}{\hat{2}}
\newcommand{\hap}{\hat{p}}
\newcommand{\haP}{\hat{P}}
\newcommand{\hal}{\hat{l}}

\newcommand{\etaorth}{\eta^{\perp}}
\newcommand{\etapar}{\eta^{\parallel}}

\def\mbf#1{\mathchoice{\hbox{\boldmath $\displaystyle #1$}}
        {\hbox{\boldmath $\textstyle #1$}}
        {\hbox{\boldmath $\scriptstyle #1$}}
        {\hbox{\boldmath $\scriptscriptstyle #1$}}}

\def\a{\alpha}      
\def\b{\beta}       
\def\c{\gamma}  \def\G{\Gamma}  \def\cdt{\dot\gamma}
\def\d{\delta}  \def\D{\Delta}  \def\ddt{\dot\delta}
\def\e{\epsilon}        \def\vare{\varepsilon}
\def\f{\phi}    \def\F{\Phi}    \def\vvf{\f}
\def\h{\eta}
\def\k{\kappa}
\def\l{\lambda} \def\L{\Lambda}
\def\m{\mu} \def\n{\nu}
\def\o{\omega}
\def\p{\pi} \def\P{\Pi}
\def\r{\rho}
\def\s{\sigma}  \def\S{\Sigma}
\def\t{\tau}
\def\th{\theta} \def\Th{\Theta} \def\vth{\vartheta}
\def\X{\Xeta}
\def\z{\zeta}
\font\mybb=msbm10 at 12pt
\def\bb#1{\hbox{\mybb#1}}

\font\myBB=msbm10 at 18pt
\def\BB#1{\hbox{\myBB#1}}

\begin{document}

\begin{flushright}

QMUL-PH-10-11
\end{flushright}

\vspace{20pt}

\begin{center}

%{\Large \bf Generalised Unitarity in Six Dimensions   }

{\Large \bf One-loop Amplitudes in Six-Dimensional}
\\
\vspace{0.3cm}
{\Large \bf  (1,1) Theories  from
Generalised Unitarity   }

\vspace{32pt}

{\mbox {\bf Andreas Brandhuber, Dimitrios Korres,  Daniel Koschade and Gabriele Travaglini}}%
\footnote{
{\sffamily \{\tt a.brandhuber, d.korres, d.koschade, g.travaglini\}@qmul.ac.uk }}

{\em Centre for Research in String Theory\\
Department of Physics\\
Queen Mary, University of London\\
Mile End Road, London, E1 4NS\\
United Kingdom
 }

\vspace{30pt} {\bf Abstract}
\end{center}

\noindent
Recently, the  spinor helicity formalism and on-shell superspace were developed for six-dimensional gauge theories with (1,1) supersymmetry. We combine these two techniques with (generalised) unitarity, which is a powerful technique to calculate scattering amplitudes in any massless theory. As an application  we calculate one-loop superamplitudes with four and five external particles in the (1,1) theory and perform several consistency checks on our results.

\noindent

\setcounter{page}{0}
\thispagestyle{empty}
\newpage

%%%%%%%%%%%%%%%%%%%%%%%%%%%%%%%%%%%%%%%%%%%%%%%%%%%%%%%%%%%%%

\section{Introduction  }

There are several reasons why it is interesting to consider scattering amplitudes in six-dimensional theories. 
Firstly, there is a powerful spinor helicity formalism, introduced in \cite{cheung} and further discussed in \cite{Boels:2009bv} for arbitrary dimensions, which allows one to express scattering amplitudes in a rather compact form. An important difference with respect to the four-dimensional  world is that physical states are no longer labeled by their helicity, but carry indices of the little group 
$\mathsf{SU}(2) \times \mathsf{SU} (2)$ of a massless particle. 
As a consequence, states in a particular little group representation can be rotated into each  other, and hence, at a fixed number of external legs, all scattering amplitudes for different external states are collected into a single object, transforming covariantly under the little group.
In \cite{cheung}, an expression for the three-point gluon amplitude in Yang-Mills theory was  obtained, and used to derive tree-level four- and five-point amplitudes using on-shell recursion relations \cite{bcf,bcfw}. 

Particularly interesting are  the maximally supersymmetric theories in six dimensions, with (1,1)  and
(2,0) supersymmetry, which arise as the low-energy effective field theories on fivebranes in string/M-theory and upon compactification on a two-torus reduce to $\cN=4$ super Yang-Mills (SYM) in four dimensions. The scattering superamplitudes in  the (1,1) theory have been studied in \cite{siegel} (see also \cite{Huang:2010rn}), using supersymmetric on-shell recursion relations \cite{ahck,bhtrec}. In particular, the three-, four- and five-point superamplitudes at tree-level have been derived, as well as the the one-loop four-point superamplitude, using the unitarity-based approach of \cite{bddk,fusing}.  Some generalisations to (2,0) theories in  six dimensions have been considered in \cite{Huang:2010rn}.

Six-dimensional tree-level amplitudes take a rather compact form, which can be fed into unitarity \cite{bddk,fusing} and generalised unitarity cuts \cite{Bern:1997sc, bcfgen} to generate loop amplitudes. Originally the unitarity methods and their generalisations were formulated in four dimensions but they apply in principle in any number of dimensions, which is also often exploited in calculations
of QCD  amplitudes in dimensional regularisation (see e.g. \cite{vanneerven,Bern:1995db,Brandhuber:2005jw,Anastasiou:2006jv}). First applications of
unitarity to one-loop four-point amplitudes in six-dimensional (1,1) theories appeared in \cite{siegel} and more recently in six-dimensional Yang-Mills in \cite{Bern:2010qa}, where also higher-loop four-point amplitudes in the (1,1) theory were computed.

Gauge theories in more than four dimensions are usually non-renormalisable,  but at least for the maximally supersymmetric examples their known embedding into string theory as low-energy theories living on D-branes or M-branes guarantees the existence of a UV completion. In particular, it is known that the (1,1) supersymmetric gauge theory in six dimensions is  finite up to two loops \cite{Howe:1983jm}. Furthermore, infrared divergences are absent in more than four dimensions, and hence all amplitudes in the (1,1) theory are expected to be finite up to two-loop order and can be calculated without regularisation. 

An additional motivation to study higher-dimensional theories stems from the fact that 
QCD amplitudes in dimensional regularisation naturally give rise to integral functions in higher dimensions, in particular
$D=6$ and $D=8$ \cite{vanneerven,Bern:1995db}. These integrals are related to finite, rational terms or  terms that vanish in the four-dimensional limit. Furthermore, there exists a mysterious dimension shift relation between MHV one-loop amplitudes in the maximally supersymmetric gauge theory in eight dimension (with four-dimensional external momenta) and the finite same-helicity one-loop gluon amplitude in pure Yang-Mills in four dimensions \cite{Bern:1996ja}.

In this paper, we calculate four- and five-point superamplitudes in the maximally supersymmetric (1,1) theory using two-particle as well as quadruple cuts at one loop.  In particular,  we show that the five-point superamplitude can be expressed in terms of just a linear pentagon integral in six dimensions, which can be further reduced in terms of  scalar pentagon and box functions.  Because of the non-chiral nature of the (1,1) on-shell superspace, this superamplitude contains all possible component amplitudes with five particles, in contradistinction with the four-dimensional case where one has to distinguish MHV and anti-MHV helicity configurations.

The rest of this paper is organised as follows. In Section 2, we review the spinor helicity formalism in six dimensions, and the on-shell (1,1) superspace, which is used to describe superamplitudes in the (1,1) theory. In Section 3 we collect the expressions for the simplest tree-level amplitudes.  These are used in Section 4 and 5 for our calculations of one-loop amplitudes using (generalised) unitarity. In Section 4 we illustrate the method
by  rederiving the four-point superamplitude using two-particle cuts as well as quadruple cuts. Next, in Section 5 we present in detail the derivation of the five-point superamplitude in the (1,1) theory from quadruple cuts. Finally, we perform several consistency checks of our result using dimensional reduction to four dimensions in order to compare with the corresponding amplitudes in $\cN=4$ SYM. We also test some of the soft limits. 

\setcounter{footnote}{0}

\section{Background} 
We begin this section by briefly reviewing the six-dimensional spinor helicity formalism developed in \cite{cheung},  
which is required to present Yang-Mills scattering amplitudes in a compact form. We then discuss 
the on-shell (1,1) superspace description of amplitudes in maximally supersymmetric Yang-Mills 
which was introduced  in \cite{siegel}. 

\subsection{Spinor helicity formalism in six dimensions}

The key observation for a compact formulation of amplitudes in six-dimensional gauge and gravity theories is that, similarly to four dimensions, null momenta in six dimensions can be conveniently presented in a spinor helicity formalism, introduced in \cite{cheung}. 
Firstly, one rewrites vectors of the Lorentz group $\mathsf{SO}(1,5)$ as antisymmetric $\mathsf{SU}(4)$ matrices 
\beq
p^{AB}\ := \ p^\mu \tilde\sigma_\mu^{AB}\ , 
\eeq
 using the appropriate  Clebsch-Gordan symbols $\tilde\sigma_\mu^{AB}$, where 
$A, B  =1, \ldots , 4$ are fundamental indices of $\mathsf{SU}(4)$.  
One can similarly introduce%
\footnote{Our notation and conventions are outlined in Appendix \ref{appA}.}  
\beq
p_{AB} \ := \ {1\over 2} \eps_{ABCD} p^{CD}  := p^\mu \sigma_{\mu, A B}
\ , 
\eeq
with 
$\sigma_{\mu, AB } := ( 1/2 )\eps_{ABCD} \tsigma_\mu^{CD}$. When $p^2 =0$, it is natural to recast $p^{AB}$ and $p_{AB}$ as the product of two spinors as \cite{cheung}
\beqa
\label{p} 
p^{AB} & = &  \l^{A a} \l^B_a
\ , 
\\ \nonumber 
p_{AB} & = & {\lt}_{A}^{\da} {\tilde \lambda}_{B \da}
\ . 
\eeqa
Here  $a=1,2$ and $\da=1,2$ are indices of the little group%
\footnote{Or $\mathsf{SL}(2,\mathbb{C})  \times \mathsf{SL}(2,\mathbb{C})$, if we complexify  spacetime.} 
 $\mathsf{SO}(4) \simeq \mathsf{SU}(2) \times \mathsf{SU}(2)$, which are contracted with the usual invariant tensors $\eps_{ab}$ and $\eps_{\da \db}$. The expression for $p$ given in \eqref{p} automatically ensures that $p$ is a null vector, since
\beq
p^2 = -{1\over 8} \eps_{ABCD} \l^A _a \l^B_b  \l^C_c \l^D_d \eps^{ab} \eps^{cd} \ = \ 0 
\ . 
\eeq
The dot product of two null vectors $p_i$ and $p_j$ can also be conveniently written using spinors as 
\beq
p_i\cdot p_j \ = \ -{1\over 4} p^{AB}_i p_{j; AB} \ . 
\eeq
%
%In order to shorten spinor expressions it is useful to introduce a bracket notation for six-dimensional spinors. 
%Assigning $| p_a \ra := \lambda^A_a$ and $|Êp_{\da} ] := \tlambda_{A, \da}$ we can conveniently write a momentum $p_i$ as
% \be
% p_i^{A B} = | i^a \ra \la i_a | \ , \; \; \; \; p_{i, AB} = | i^{\da} ] [ i_{\da} | 
 %\ee
% where the $\mathsf{SU}(4)$ Lorentz indices $A,B$ are understood implicitly. From this 
%
Lorentz invariant contractions of two spinors are expressed as
 \be
 \la i_a |Êj_{\da} ] := \lambda_{i,a}^{A } \tlambda_{j, A \da} = \tlambda_{j, A \da} \lambda_{i,a}^{A }  =: [ j_{\da} |Êi_a \ra \ .
\ee
Further  Lorentz-invariant combinations can be constructed from four spinors using the $\mathsf{SU}(4)$ invariant $\e$ tensor, as 
\beqa 
\la 1_a \, 2_b \, 3_c \, 4_d \ra &:= &\eps_{ABCD}\l_{1, a}^A \l_{2, b}^B\l_{3, c}^C\l_{4, d}^D \ , 
\\ \nonumber 
[  1_{\da} \, 2_{\db} \, 3_{\dc} \, 4_{\dd} ] &:= &\eps^{ABCD} \lt_{1, A \da} \lt_{2, B \db}\lt_{3, C \dc} \lt_{4, D \dd}  \ . 
\eeqa
This notation may be used to express compactly strings of six-dimensional momenta contracted with Dirac matrices, such as
\begin{alignat}{1}
\la i_a |Ê\hap_1 \hap_2 \dots \hap_{2n+1} |Êj_b \ra & := \lambda_{i, a}^{A_1} \; p_{1, A_1 A_2} \; p_2^{A_2 A_3} \dots p_{2n+1, A_{2n+1} A_{2n+2}} \; \lambda_{j, b}^{A_{2n+2}} \ ,  \\
\la i_a |ÊÊ\hap_1 \hap_2 \dots \hap_{2n} | j_{\da } ] & := \lambda_{i, a}^{A_1} \; p_{1, A_1 A_2} \; p_2^{A_2 A_3} \dots  p_{2n}^{A_{2n} A_{2n + 1}} \; \tlambda_{j, A_{2n+1} \db} \nonumber \ .
\end{alignat}
Having discussed momenta, we now consider polarisation states of particles. In four dimensions, these are associated to the notion of helicity. In six dimensions, physical states, and hence their  wavefunctions, transform according to representations of the  little group, and therefore carry $\mathsf{SU}(2)\times \mathsf{SU}(2)$ indices \cite{cheung}. In particular, for gluons  of momentum $p$ defined as in \eqref{p} one has 
%\begin{alignat}{1}
%\epsilon^{\mu}_{ a \da} (p, q) & := \frac{1}{2} \frac{\la p_a |Ê\sigma^{\mu} | q_b \ra }{\la q_b |Êp^{\da} ]}  = \frac{1}{2} %\frac{ [ q_{\db} |Ê\tsigma^{\mu} | p_{\da} ] }{\la p^a |Êq_{\db} ]} \ .
%\end{alignat}
\beq
\eps_{ a \da}^{AB} :=   \l_a^{[A} \eta_b^{B]} \langle \eta_b | \lt^{\da} ]^{-1}
\ , 
\eeq 
or alternatively 
\beq
\eps_{ a \da; AB} :=     \langle \l^a | \tilde{\eta}_{\db} ]^{-1}    \tilde{\eta}_{\dot{b} [ A}   \lt_{\da B] }
\ . 
\eeq 
Here, $\eta$ and $\tilde\eta$ are reference spinors, and the denominator is defined to be the inverse of the matrices $\la q^b |Êp_{\da}Ê] $ and $\la p_a | q^{\db} ]$, respectively.%
\footnote{The reference spinors are chosen such that the  matrices $\la q^b |Êp_{\da}Ê] $ and $\la p_a | q^{\db} ]$ are nonsingular.}

It is amusing to make contact between six-dimensional spinors  and  momentum twistors \cite{hodges},  employed recently to describe amplitudes in four-dimensional conformal theories. 
There, one describes a point in (conformally compactified) Minkowski space as a six-dimensional null vector $X$, i.e.~one satisfying $\eta_{ij} X^i X^j = 0$, with 
$\eta = {\rm diag}(+---; +-)$.  The conformal group $\mathsf{SO}(2,4)$ acts linearly on the $X$ variables, and plays the role of the Lorentz group $\mathsf{SO}(1,5)$ acting on  our six-dimensional momenta $p$. Furthermore,  in contradistinction with the null six-dimensional momenta,  the coordinate $X$ are defined only up to nonvanishing rescalings.
For (cyclically ordered) four-dimensional region momenta $x_i$, one defines the corresponding six-dimensional null $X_i$ as 
$X_i \ = \l_i \wedge \l_{i+1}$ , $X_j  =\l_j \wedge \l_{j+1}$,
and 
$ X_i \cdot X_j \ = \ \la i \, i+1\,  j\,  j+1 \ra$. 
\subsection{(1,1) on-shell superspace}

We will now review the on-shell superspace description of $(1,1)$ theories introduced in \cite{siegel}. This construction is inspired by the covariant on-shell superspace formalism for four-dimensional $\cN=4$ SYM introduced by Nair in \cite{Nair}. 
 In the latter case, the $\mathcal{N}=4$ algebra can be represented on shell as
\be
\{ q^A_{\alpha}, \tilde{q}_{B \dot{\alpha}} \} = \delta^A_B \lambda_{\alpha} \tlambda_{\dot{\alpha}}\ , 
\ee
where $A, B$ are $\mathsf{SU}(4)$ $R$-symmetry indices and $\alpha, \dot{\alpha}$ are the usual $\mathsf{SU}(2)$ spinor indices in four dimensions. The supercharge $q$ can be decomposed along two independent directions $\lambda$ and $\mu$  as
\be
q^A_{\alpha} = \lambda_{\alpha} q^A_{(1)}  + \mu_{\alpha} q^A_{(2)} \ , 
\ee 
where $\langle \lambda \mu \rangle \neq 0$. A similar decomposition is performed for $\tilde{q}$. 
One can then easily see that the charges $q_{(2)}$ and $\tilde{q}_{(2)}$ anticommute among themselves and with the other generators, and can therefore be set to zero. 
The supersymmetry algebra becomes
\be
\{ q^A_{(1)} , \tilde{q}_{(1) B} \} = \delta^A_B  \ .
\ee
Setting $q^A_{(1)}  = q^A$ and $\tilde{q}_{(1) B} = \tq_B$, the Clifford algebra can be naturally realised in terms of Grassmann variables $\eta^A$, as
\be
\label{qqq}
q^A = \eta^A \ , \qquad \qquad  \tq_A = \frac{\partial}{\partial \eta^A} \ .
\ee
Note that this representation of the algebra is chiral. One could have chosen an anti-chiral representation, where  the roles of $q$ and $\tilde{q}$ in \eqref{qqq} are interchanged. 

One can apply similar ideas to the case of the $\mathcal{N}=(1,1)$ superspace of the six-dimensional SYM theory. However, for this on-shell space the chiral and anti-chiral components do not decouple. To see this we start with the algebra
\begin{alignat}{1}
\{ q^{A I}, q^{B J} \} &  = p^{A B} \epsilon^{I J} \ , 
\\
\{ \tq_{A I'}, \tq_{B J'} \} &  = p_{A B} \epsilon_{I' J'}\ , 
 \nonumber 
\end{alignat}
where $A, B$ are the $\mathsf{SU}(4)$ Lorentz index and $I, J$ and $I', J'$ are indices of the $R$-symmetry group $\mathsf{SU}(2) \times \mathsf{SU}(2)$. As before, we decompose the supercharges as
\begin{alignat}{1} \label{6d-supercharge-decomposition}
q^{A I} & = \lambda^{A a} q_{(1) a}^I + \mu^{A a} q_{(2) a}^I \ , \\
\tq_{B I'} & = \tlambda_{B}^{\db} \tq_{(1) \db I'} + \tilde{\mu}_{B}^{\db} \tq_{(2) \db I'} \ ,  \nonumberâ
\end{alignat}
with ${\rm det} (\lambda^{A a} \tilde{\mu}_{A}^{\da} ) \neq 0$ and  ${\rm det}( \mu^{A a} \tlambda_{A}^{\da}) \neq 0$. 
Multiplying  the supercharges in (\ref{6d-supercharge-decomposition}) by  $\tlambda_{A \da}$ and $\lambda^B_b$, respectively, and summing over the {\sf SU}(4) indices, one finds that 
\begin{alignat}{1}
\{ q^I_{(2) a}, q^J_{(2) b} \} & = 0  \  , \\
\{ \tq_{(2) \da I'}, \tq_{(2) \db J'} \} & = 0 \ . \nonumber 
\end{alignat}
\begin{figure}[t]
\begin{center}
\includegraphics[width=7cm]{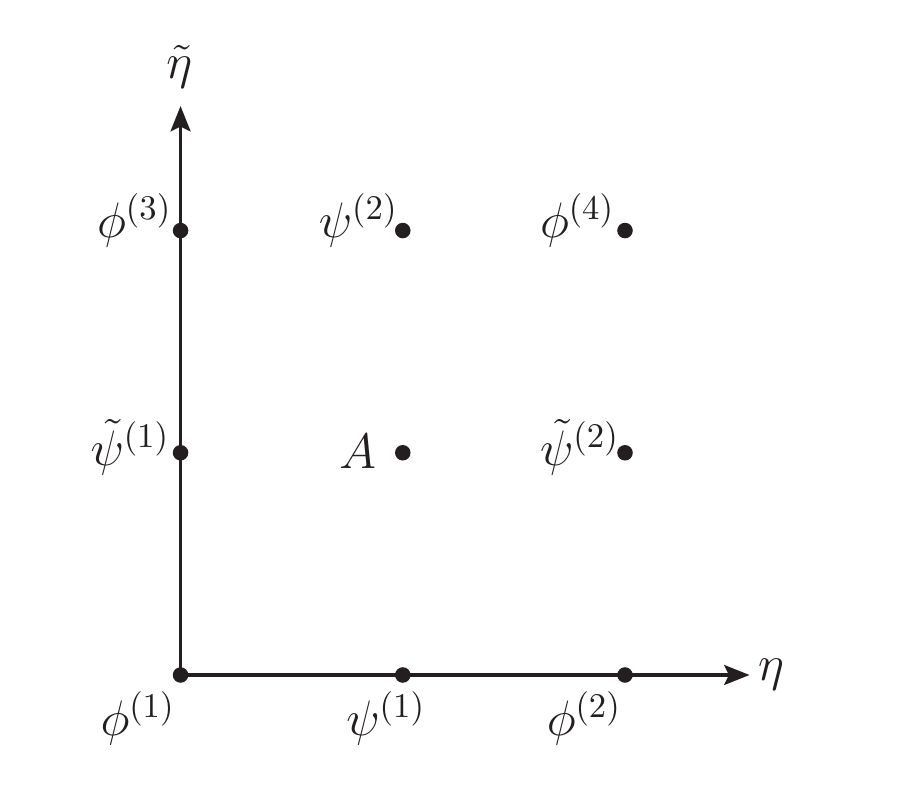}
\caption{\it The component fields of the  (1,1) superfield given in \eqref{6d-superfield}.}
\label{superspace-diagram}
\end{center}
\end{figure}
One can thus set all the $q_{(2) }$  and $\tilde{q}_{(2) }$ charges equal to zero, so that  $q^{A I} = \lambda^{A a} q^I_{(1) a}$. 
The supersymmetry algebra then yields, 
\beqa
\label{blam}
\{ q^{I}_{(1)a }, q^J_{(1) b} \} &=& \eps_{ab} \epsilon^{I J}
\ , \nonumber \\
\{ \tq_{(1) I'\da}, \tq_{(1) J' \db} \} &=& \eps_{\da\db}\epsilon_{I' J'} \ .
\eeqa
The realisation of \eqref{blam}   in terms of anticommuting Grassmann variables is 
\be \label{full-6d-supercharges}
q^{A I} = \lambda^{A a} \eta^I_a \, , \quad 
\qquad \tq_{A I'} = \tlambda_A^{\da} \teta_{I' \da} \ . 
\ee
In contrast to the four-dimensional $\mathcal{N}\!=\!4$ SYM theory, the $\mathcal{N}=(1,1)$ on-shell superspace in six dimensions carries chiral and anti-chiral components. The field strength of the six-dimensional SYM theory transforms under the little group
$\mathsf{SU}(2) \times \mathsf{SU}(2)$ and therefore carries both indices $a$ and $\da$. Hence, one needs both $\eta_a$ and $\teta_{\da}$ to describe all helicity states  in this theory.   

In order to describe only the physical components of the full six-dimensional SYM theory, one needs to truncate half of the superspace charges in  \eqref{full-6d-supercharges} \cite{siegel}. This is performed by contracting the $R$-symmetry indices with fixed two-component (harmonic) vectors, which effectively reduce the number of supercharges by a factor of two. The resulting  truncated supersymmetry generators  are then  \cite{siegel}
\be \label{trun-6d-supercharges}
q^{A } = \lambda^{A a} \eta_a \, , 
\qquad \quad \tq_{A } = \tlambda_A^{\da} \teta_{\da}
\ . 
\ee
Using this on-shell superspace, one can neatly package all states of the theory into a six-dimensional analogue of  Nair's superfield  \cite{Nair}, 
\beqa
\label{6d-superfield}
%\begin{alignat}{1}
\Phi( p; \eta, \tilde{\eta})   &= & \phi^{(1)} + \psi_a^{(1)} \eta^a + \tpsi_{\da}^{(1)} \teta^{\da} + \phi^{(2)} \eta^a \eta_a + A_{a \da} \eta^a \teta^{\da} + \phi^{(3)} \teta^{\da} \teta_{\da} \\ \nonumber 
& + &\psi_a^{(2)} \eta^a \teta^{\da} \teta_{\da} + \tpsi^{(2)}_{\da} \teta^{\da} \eta^a \eta_a + \phi^{(4)} \eta^a \eta_a \teta^{\da} \teta_{\da}
\ . 
\eeqa
Here  $\phi^{(i)}(p) $, $i=1, \ldots, 4$ are four scalar fields, $\psi^{(l)}(p)$ and $\tilde\psi^{(l)}(p)$, $l=1,2$ are fermion fields and finally $A_{a \da}(p)$ contains the gluons. Upon reduction to four dimensions, $A_{a \da}$ provides, in addition to   gluons of positive and negative (four-dimensional) helicity, the two remaining scalar fields needed to obtain the matter content of $\cN=4$ super Yang-Mills.%
\footnote{More details on reduction to four dimensions are provided in Section \ref{4dconcheck}.} 
A pictorial representation of the states in the (1,1) supermultiplet is given in Figure \ref{superspace-diagram}.

\section{Tree-level amplitudes and their properties}

\subsection{Three-point amplitude}

The smallest amplitude one encounters is the three-point amplitude. In four dimensions, and for  real kinematics, three-point amplitudes   vanish  because $p_i \cdot p_j = 0$ for any of the three particles' momenta,  but are non-vanishing upon spacetime complexification \cite{bcf,bcfw}. 
In the six-dimensional spinor helicity formalism,  the special three-point kinematics induces the constraint 
${\rm det} \la i_a |Êj_{\da} ] = 0$, $i,j=1,2,3$.  
This allows one to write (see Appendix \ref{appA})
\be
\la i_{a} | j_{\dot{b}} ]= (-)^{\cP_{ij}} u_{ia} \tilde{u}_{j\dot{b}}
\ , 
\ee
where we choose $(-)^{\cP_{ij}} = +1$ for $(i,j) =(1,2), (2,3), (3,1)$ and $-1$ for   $(i,j) =(2,1), (3,2), (1,3)$. 
One can also introduce the spinors $w_a$ and $\tw_{\da}$ \cite{cheung}, defined as the inverse of $u_a$ and $\tu_{\da}$, \be
u_{a} w_{b} - u_{b} w_{a} := \epsilon_{a b}  \; \; \Leftrightarrow \;   \;  u^a w_a := - u_a w^a := 1 \, . 
\ee
As stressed in \cite{cheung} the $w_i$ spinors are not uniquely specified. Momentum conservation suggests a further constraint that may imposed in order to reduce this redundancy. This is used in various calculations throughout the present work. Specifically, for a generic three-point amplitude it is assumed that 
\beq \label{wconstraint}
|w_{1}\cdot1\rangle+|w_{2}\cdot2\rangle+|w_{3}\cdot3\rangle=0
\ .
\eeq
One may then express the three-point tree-level amplitude for six-dimensional Yang-Mills theory as \cite{cheung}
\be \label{tree-3pt}
A_{3;0} (1_{a \da}, 2_{b \db}, 3_{c \dc} ) = i \Gamma_{a b c} \tG_{\da \db \dc}
\ , 
\ee
where the tensors $\Gamma$ and $\tG$ are given by
\begin{alignat}{1}
\Gamma_{a b c} & =  u_{1 a} u_{2 b} w_{3 c} + u_{1 a} w_{2  b} u_{3 c} + w_{1 a} u_{2 b} u_{3 c},  \\ 
\tG_{\da \db \dc} & = \tu_{1 \da} \tu_{2 \db} \tw_{3 \dc} + \tu_{1 \da} \tw_{2 \db} \tu_{3 \dc} + \tw_{1 \da} \tu_{2 \db} \tu_{3 \dc} \ .  \nonumber  
\end{alignat}
As recently shown in \cite{siegel}, this result can be combined with the $\mathcal{N}=(1,1)$ on-shell superspace in six dimensions. The corresponding three-point tree-level superamplitude  takes the simple form \cite{siegel}
\be \label{tree-3pt-super}
A_{3;0} (1_{a \da}, 2_{b \db}, 3_{c \dc} ) \ = \ i\, \delta (Q^A) \delta (\tQ_A) \delta (Q^B) \delta (\tQ_B) \delta (W) \delta (\tW)
\ . 
\ee
Here we have introduced  the $\mathcal{N}=(1,1)$ supercharges for  the external states, 
\begin{eqnarray} \label{susygen}
Q^{A}\, := \, \sum_{i=1}^n q_i^A=\sum_{i=1}^n\lambda_{i}^{A a}\eta_{i a}{}^{} \ , \qquad && \tilde{Q}_{A}\, := \, \sum_{i=1}^n \tq_{iA}=\sum_{i=1}^n\tilde{\lambda}_{i A}^{\dot{a}} \tilde{\eta}_{i \dot{a}}
\ 
\end{eqnarray}
(with $n=3$ in the three-point amplitude we are considering in this section). 
The quantities $W, \tW$ appear only in the special three-point kinematics case, and are given by 
\be 
\label{w}
W\, := \, \sum_{i=1}^3 w_i^a \eta_{i a}\,  , \qquad  \tW \, :=\, \sum_{i=1}^3 \tw_i^{\da} \teta_{i \da} \ .
\ee
In Appendix \ref{appD} we give an explicit proof of the (non manifest) invariance of the three-point superamplitude under supersymmetry transformations, and hence of the fact that the total supermomentum $Q^A = \sum_i q_i^A$  is conserved. 

\subsection{Four-point amplitude}

The four-point tree-level amplitude in six dimensions is given by 
\be \label{tree-4pt}
A_{4;0} (1_{a \da}, 2_{b \db}, 3_{c \dc}, 4_{d \dd} ) = -\frac{i}{s t} \la 1_a 2_b 3_c 4_d \ra [ 1_{\da} 2_{\db} 3_{\dc} 4_{\dd} ]
\ , 
\ee
and was derived by using a six-dimensional version \cite{cheung} 
of the BCFW recursion relations \cite{bcf,bcfw}. 
The corresponding  $\mathcal{N}=(1,1)$ superamplitude is \cite{siegel} 
\be 
\label{tree-4pt-super}
A_{4;0} (1, \ldots , 4) = - \frac{i}{s t} \delta^4 (Q) \delta^4 (\tQ) \ , 
\ee
where the $ (1,1)$ supercharges are defined in \eqref{susygen}. 
In (\ref{tree-4pt-super}) we follow \cite{siegel} and introduce the fermionic $\delta$-functions which enforce supermomentum conservation as
\beqa
\label{sumom}
\delta^4 (Q) \delta^4 (\tQ ) &=& \frac{1}{4!}  \epsilon_{A B C D} \delta (Q^A) \delta (Q^B) \delta (Q^C) \delta (Q^D) \nonumber \\ &\times &   \frac{1}{4!}   \epsilon^{A^\prime B^\prime C^\prime D^\prime} 
\delta (\tilde{Q}_{A'}) 
\delta (\tilde{Q}_{B'}) 
\delta (\tilde{Q}_{C'}) 
\delta (\tilde{Q}_{D'}) 
\nonumber \\ 
&:=& \delta^8 (Q)
\ .
\eeqa
Hence, a $\delta^4 (Q)$ sets $Q^A =0$ whereas the $\delta^4 (\tQ)$ sets $\tQ_A =0$. 

\subsection{Five-point amplitude}

The five-point tree-level amplitude was  derived in \cite{cheung} using recursion relations, and is equal to%
\footnote{
In Appendix \ref{appE} the five-point amplitude \eqref{5pttree} is reduced to four dimensions and found to be in agreement with the expected Parke-Taylor expression.  }  
\beq \label{5pttree}
A_{5;0} (1_{a \da}, 2_{b \db}, 3_{c \dc}, 4_{d \dd}, 5_{e \de} ) \ = \  \frac{i}{s_{12} s_{23} s_{34} s_{45} s_{51}} \left( \mathcal{A}_{a \da b \db c \dc d \dd e \de} + \mathcal{D}_{a \da b \db c \dc d \dd e \de} \right)
\eeq
where the two tensors $\mathcal{A}$ and $\mathcal{D}$ are given by 
\be
 \mathcal{A}_{a \da b \db c \dc d \dd e \de} = \la 1_a | \hap_2 \hap_3 \hap_4 \hap_5 | 1_{\da} ] \la 2_b 3_c 4_d 5_e \ra [ 2_{\db} 3_{\dc} 4_{\dd} 5_{\de}Ê] + \mbox{cyclic permutations} \ ,
\ee
and
\begin{alignat}{1}
2 \mathcal{D}_{a \da b \db c \dc d \dd e \de} & =  \la 1_a | (2 \cdot \tD_2)_{\db} ] \la 2_b 3_c 4_d 5_e \ra [ 1_{\da} 3_{\dc} 4_{\dd} 5_{\de} ] + \la 3_c | (4 \cdot \tD_4)_{\dd} ] \la 1_a 2_b 4_d 5_e \ra [ 1_{\da} 2_{\db} 3_{\dc} 5_{\de} ]  \nonumber \\
& + \la 4_d | (5 \cdot \tD_5)_{\de} ] \la 1_a 2_b 3_c 4_d \ra [ 1_{\da} 2_{\db} 3_{\dc} 4_{\dd} ] - \la 3_c | (5 \cdot \tD_5)_{\de} ] \la 1_a 2_b 4_d 5_e \ra [ 1_{\da} 2_{\db} 3_{\dc} 4_{\dd} ] \nonumber \\
& - [ 1_{\da} |Ê(2 \cdot \Delta_2)_b \ra \la 1_a 3_c 4_d 5_e \ra [ 2_{\db} 3_{\dc} 4_{\dd} 5_{\de} ] - [ 3_{\dc} |Ê(4 \cdot \Delta_4)_d \ra \la 1_a 2_b 3_c 5_e \ra [ 1_{\da} 2_{\db} 4_{\dd} 5_{\de} ] \nonumber \\
& - [ 4_{\dd} |Ê(5 \cdot \Delta_5)_e \ra \la 1_a 2_b 3_c 4_d \ra [ 1_{\da} 2_{\db} 3_{\dc} 5_{\de} ] +  [ 3_{\dc} |Ê(5 \cdot \Delta_5)_e \ra \la 1_a 2_b 3_c 4_d \ra [ 1_{\da} 2_{\db} 4_{\dd} 5_{\de} ] \ .
\end{alignat}
Here, the spinor matrices $\Delta$ and $\tD$ are defined by
\be
\label{ab}
\Delta_1 = \la 1 |Ê\hap_2 \hap_3 \hap_4 - \hap_4 \hap_3 \hap_2 | 1 \ra, \; \; \; \; \tD_1 = [ 1 | Ê\hap_2 \hap_3 \hap_4 - \hap_4 \hap_3 \hap_2 | 1 ] 
\ , 
\ee
where the other quantities $\Delta_i, \tD_i$ are generated by taking cyclic permutation on \eqref{ab}.  The contraction between a $\Delta_i$ and the corresponding spinor $\lambda_i^{A a}$ is given by   $\la 1_a | (2 \cdot \tD_2)_{\db} ]  = \lambda^A_{1 a} \tlambda_{2 A}^{\da'} [ 2_{\da'} | Ê\hap_3 \hap_4 \hap_5 - \hap_5 \hap_4 \hap_3 | 2_{\db} ] $.

The five-point superamplitude in the $\mathcal{N}=(1,1)$ on-shell superspace can also be calculated in a recursive fashion. It takes the form \cite{siegel}
\begin{alignat}{1}
A_{5;0}  \ = \ & i\, \frac{\delta^4 (Q) \delta^4 (\tQ)}{s_{12} s_{23} s_{34} s_{45} s_{51} } \bigg[  \\
& + \frac{3}{10} q_1^A \big[ (\hap_2 \hap_3 \hap_4 \hap_5) - (\hap_2 \hap_5 \hap_4 \hap_3) \big]_A^B \tq_{2 B} +  \frac{3}{10} \tq_{1 A} \big[ (\hap_2 \hap_3 \hap_4 \hap_5) - (\hap_2 \hap_5 \hap_4 \hap_3) \big]^A_B q_2^B \nonumber \\
& + \frac{1}{10} q_3^A \big[ (\hap_5 \hap_1 \hap_2 \hap_3) - (\hap_5 \hap_3 \hap_2 \hap_1) \big]_A^B  \tq_{5 B} + \frac{1}{10} \tq_{3 A} \big[ (\hap_5 \hap_1 \hap_2 \hap_3) - (\hap_5 \hap_3 \hap_2 \hap_1) \big]^A_B  q_{5 }^B \nonumber \\
& + q_1^A (\hap_2 \hap_3 \hap_4 \hap_5)_A^B \tq_{1 B}+ \mbox{cyclic permutations} \bigg] 
\ , \nonumber
\end{alignat}
where the supercharges $Q$ and $\tilde{Q}$ are defined in \eqref{susygen}.

\section{One-loop four-point amplitude} 
In this section we calculate the four-point one-loop amplitude using two-particle and four-particle cuts. As expected, we find that the one-loop amplitude is proportional to the four-point tree-level superamplitude times the corresponding integral function.

\subsection{The superamplitude from two-particle cuts\label{sub:Two-particle-cuts}}

\begin{figure}[t]
\begin{center}
\includegraphics[width=8cm]{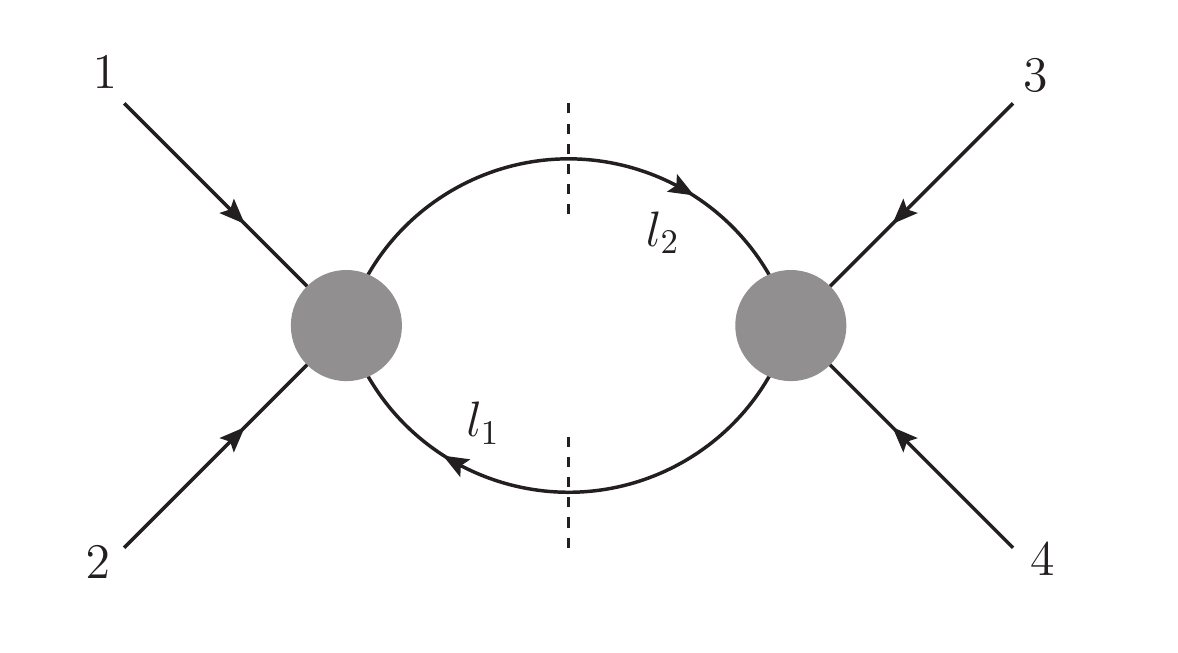}
\caption{\it Double cut in the s-channel. The two internal cut-propagators, carrying momenta $l_1$ and $l_2$ set the two four-point subamplitudes on-shell. We identify $l_1 = l$ and $l_2 = l + p_1 + p_2$.}
\label{one-loop-double-cut}
\end{center}
\end{figure}

As a warm-up exercise,  we start by rederiving  the one-loop four-point superamplitude in six dimensions using two-particle cuts. This calculation  was first {sketched} in \cite{siegel}. Here, we will perform it in some detail while setting up our notations.  We will then show how to reproduce this result using quadruple cuts. 

We begin by considering the one-loop amplitude with external momenta $p_1, \dots,  p_4$, and perform a unitarity cut in the $s$-channel, see Figure \ref{one-loop-double-cut}. %where we identify 
%\be
%l_1 = l, \; l_2 = l + p_1 + p_2
%\ . 
%\ee
The $s$-cut of the one-loop amplitude is given by\footnote{See Appendix A for our definitions of fermionic integrals.}
\begin{alignat}{1}
\label{cut1}
A_{4;1} \big|_{\mbox{\scriptsize $s$-cut}} = & \int\!\!\frac{d^6 l}{(2 \pi)^6} \delta^{+} (l_1^2) \delta^{+} (l_2^2) \, \bigg[ \prod_{i=1}^2 \int d^2 \eta_{l_i} d^2 \teta_{l_i}\,  A_{4;0}^{(L)} (l_1, 1, 2, - l_2)  \,   A_{4;0}^{(R)} (l_2, 3, 4, - l_1)  \bigg] 
\ . 
\end{alignat}
Plugging the expression \eqref{tree-4pt-super} of  the four-point  superamplitude into \eqref{cut1},
we get the following fermionic integral,
\begin{alignat}{1}
\label{gra1}
\prod_{i=1}^2 \int d^2 \eta_{l_i} d^2 \teta_{l_i} \bigg( \frac{-i}{s_L t_L} \delta^4 ( \sum_{ L} q_i ) \delta^4 ( \sum_{L} \tq_{i } ) \bigg) \bigg( \frac{-i}{s_R t_R} \delta^4 ( \sum_{ R} q_i ) \delta^4 ( \sum_{R} \tq_{i } ) \bigg)
\, , 
\end{alignat}
where the sums are over the external states of the left and right subamplitude in the cut diagram and the kinematical invariants are given by
\be
t_L = ( l_1 + p_1)^2\, , \hspace{1cm} t_R = (l_2 + p_3)^2
\, , 
\ee
and
\be
s_L = (p_1 + p_2)^2 = (p_3 + p_4)^2 = s_R = s \,.
\ee
Using supermomentum conservation we can remove the dependence of the loop-supermomenta on one side of the cut. For instance a $\delta^4 (Q_R) $ sets  $q_{l_1}^A = q_{l_2}^A + q_3^A + q_4^A$,
 which can be used in the remaining $\delta^4 (Q_L)$ to write 
\begin{alignat}{1}
& \delta^4 ( \sum_{ L} q_i) \rightarrow \delta^4 ( \sum_{\mbox{\scriptsize ext}} q_i ) \equiv  \delta^4 (Q_{\mbox{\scriptsize ext}}) \, ,  \nonumber \\
&  \delta^4 ( \sum_{L} \tq_{i } ) \rightarrow  \delta^4 ( \sum_{\mbox{\scriptsize ext}} \tq_{i } ) \equiv  \delta^4 (\tQ_{\mbox{\scriptsize ext}  })  \, .
\end{alignat}
Hence, \eqref{gra1} becomes
\be
\delta^4 (Q_{\mbox{\scriptsize ext}})  \delta^4 (\tQ_{\mbox{\scriptsize ext}  })    \prod_{i=1}^2 \int d^2 \eta_{l_i} d^2 \teta_{l_i} \delta^4 ( \sum_{ R} q_i ) \delta^4 ( \sum_{R} \tq_{i } ) \, . 
\ee
To perform the integration, we need to pick  two powers of $\eta_{l_i}$ and two powers of $\teta_{l_i}^{\da}$. Expanding the fermionic $\delta$-functions,  we find one possible term with the right powers of Grassmann variables to be
\be
\eta_{l_1 a} \eta_{l_1 b}  \eta_{l_2 c} \eta_{l_2 d} \teta_{l_1 \da} \teta_{l_1 \db}  \teta_{l_2 \dc} \teta_{l_2 \dd} \left[ \epsilon_{A B C D}  \lambda^{A a}_{l_1} \lambda^{B b}_{l_1} \lambda^{C c}_{l_2} \lambda^{D d}_{l_2} \;   \epsilon^{E F G H}  \tlambda^ {\da}_{l_1 E}  \tlambda^{ \db}_{l_1 F} \tlambda^{ \dc}_{l_2 G} \tlambda^{\dd}_{l_2 H} \right].  
\ee
Other combinations can be brought into that form by rearranging and relabeling indices. Integrating out the Grassmann variables gives
\be
\left( \epsilon_{A B C D} \lambda^{A a}_{l_1} \lambda^{B}_{l_1 a} \;  \lambda^{C b}_{l_2} \lambda^{D}_{l_2 b} \right) \; \left( \epsilon^{E F G H} \tlambda^ {\da}_{l_1 E}  \tlambda_{l_1 F \da} \;  \tlambda^{ \db}_{l_2 G} \tlambda_{l_2 H \db} \right) \, .
\ee
Hence, the two-particle cut reduces to
\begin{alignat}{1}
A_{4;1} \big|_{\mbox{\scriptsize $s$-cut}} \propto & (-1)\int\!\!\frac{d^6 l}{(2 \pi)^6} \delta^{+} (l_1^2) \delta^{+} (l_2^2) \, \bigg[  \delta^4 (Q_{\mbox{\scriptsize ext}})  \delta^4 (\tQ_{\mbox{\scriptsize ext}  }) \frac{\epsilon_{A B C D} l_1^{AB} l_2^{CD} \; \epsilon^{E F G H} l_{1 EF} l_{2 GH} }{s^2 (l_1+ p_1)^2 (l_2 +p_3)^2 }  \bigg] \, .  
\end{alignat}
Next, we use  \eqref{dotpr} to rewrite
\be 
\epsilon_{A B C D} \; p_{l_1}^{A B} p_{l_2}^{C D} \;  \epsilon^{E F G H} p_{l_1 E F} p_{l_2 G H} = 64 \left( l_1 \cdot l_2 \right)^2
\ . 
\ee
Thus  we obtain, for  the one-loop superamplitude, 
\begin{alignat}{1}
A_{4;1} \big|_{\mbox{\scriptsize $s$-cut}}
\propto & -   \delta^4 (Q_{\mbox{\scriptsize ext}})  \delta^4 (\tQ_{\mbox{\scriptsize ext}  })  \int\!\!\frac{d^6 l}{(2 \pi)^6} \delta^{+} (l_1^2) \delta^{+} (l_2^2) \, \bigg[ \frac{ 64 \left( l_1 \cdot l_2 \right)^2}{s^2 ( l_1 + p_1)^2 (l_2 + p_3)^2 } \bigg] \nonumber \\
= &  -\, 16\, \delta^4 (Q_{\mbox{\scriptsize ext}})  \delta^4 (\tQ_{\mbox{\scriptsize ext}  }) 
 \int\!\!\frac{d^6 l}{(2 \pi)^6} \delta^{+} (l_1^2) \delta^{+} (l_2^2) \, \bigg[   \frac{ 1}{  ( l_1 + p_1)^2 (l_2 + p_3)^2 } \bigg] \nonumber \\
= &  -  16 \;  i s t A_{4;0} (1, \ldots , 4) \; I_4 (s, t)\big|_{\mbox{\scriptsize $s$-cut}}
\ , 
\end{alignat}
where $A_{4;0} (1, \ldots , 4)$  is the tree-level four-point amplitude in \eqref{tree-4pt-super}, and  
\be
I_4 (s, t) =   \int\!\!\frac{d^6 l}{(2 \pi)^6}  \bigg[  \frac{1}{  l_1^2 l_2^2 ( l + p_1)^2 (l -p_4)^2 } \bigg] \, .
\ee
The $t$-channel cut is performed in the same fashion and after inspecting it we conclude that 
\be
A_{4;1}(1, \ldots, 4)  \, = \,   st\, A_{4;0} (1, \ldots , 4) \,  I_4 (s, t)
\ , 
\ee
in agreement with the result  of  \cite{siegel}.

\subsection{The superamplitude from quadruple cuts}

\begin{figure}
\begin{center}
\includegraphics[width=8cm]{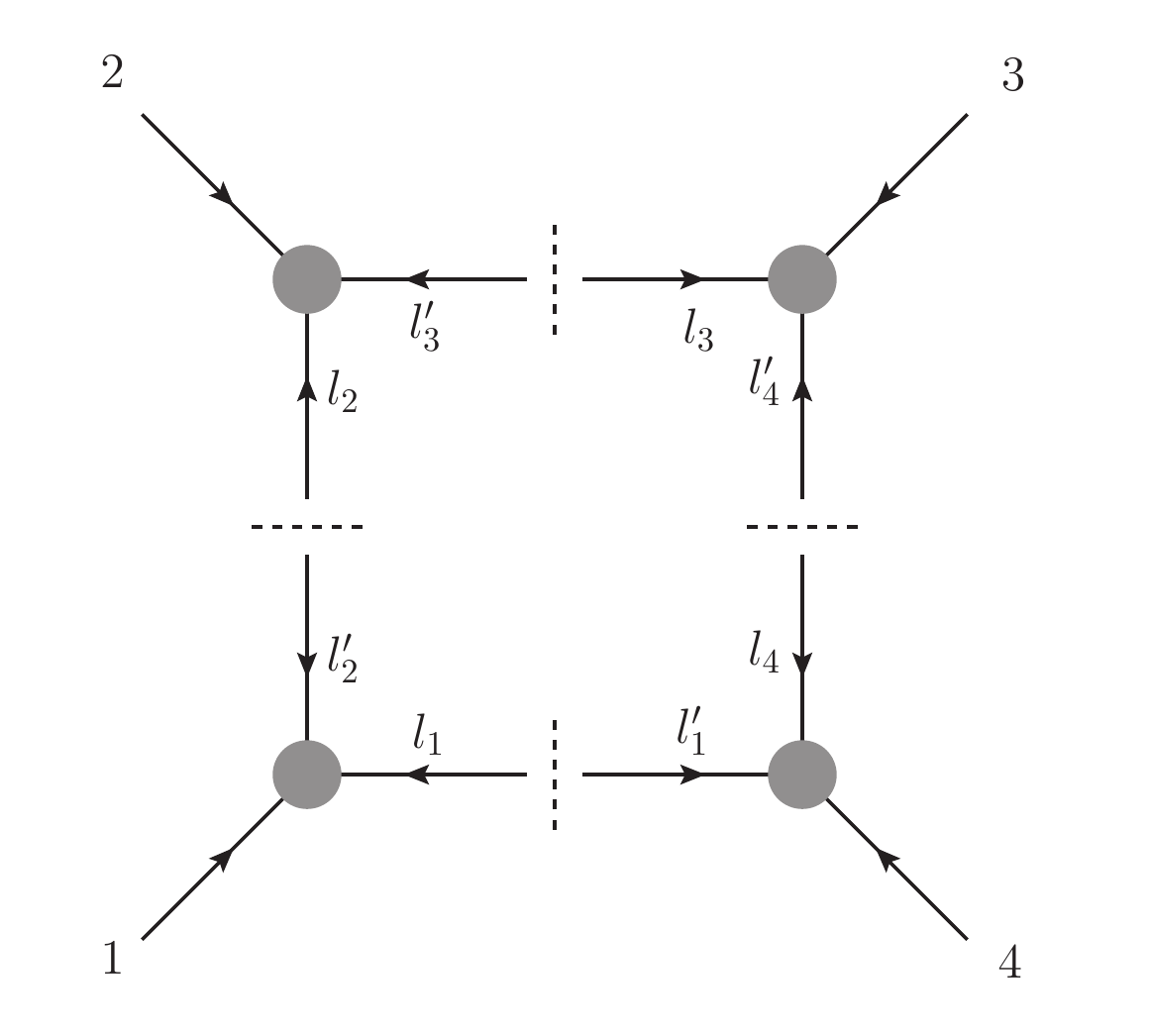}
\caption{\it The quadruple cut of a four-point superamplitude. The primed momenta $l_i'$ are defined as $l_i' := - l_i$. 
 }
\label{one-loop-quad-cut}
\end{center}
\end{figure}

We now move on to  studying the quadruple cut of  the one-loop four-point superamplitude, depicted in Figure \ref{one-loop-quad-cut}. The loop momenta are defined as 
\be
l_1 = l, \qquad 
 l_2 = l + p_1,  \qquad 
 l_3 = l + p_1 + p_2,  \qquad 
  l_4 = l - p_4\ , 
\ee
and all primed momenta $l_i'$ in Figure \ref{one-loop-quad-cut} are understood to flow in opposite direction to the $l_i$'s.  

Four three-point tree-level superamplitudes enter the quadruple cut expression.  
Uplifting the cut by replacing cut with uncut propagators, 
we obtain, for the one-loop superamplitude,
\begin{alignat}{1}
\label{poi}
A_{4;1} =  \int \frac{d^6 l}{(2 \pi)^6} \bigg[ \prod_{i=1}^4 \int & d^2 \eta_{l_i} d^2 \teta_{l_i}  \frac{1}{l_1^2} A_{3;0} (l_1,1, l_2')  \frac{1}{l_2^2} A_{3;0} (l_2, 2, l_3')  \nonumber \\
& \hspace{1cm} \times \frac{1}{l_3^2} A_{3;0} (l_3, 3, l_4')  \frac{1}{l_4^2} A_{3;0} (l_4, 4, l_1')  \bigg] \, .  
\end{alignat}
In the following we will discuss two different  but equivalent approaches  to evaluate the Grassmann integrals in \eqref{poi}.

\subsubsection{Quadruple cut as reduced two-particle cuts}

To begin with, we proceed in way similar to the case of a double cut. The idea is to integrate over two of the internal momenta, say $l_1$ and $l_3$ first, and treat $l_2$ and $l_4$ as fixed, i.e.÷ external lines. In doing so the quadruple cut splits into two four-point tree-level superamplitudes, having the same structure as in case of the BCFW construction for the four-point trevel-level superamplitude \cite{siegel}.

Let us start by focusing on the `lower' part of the diagram first. Here we have two three-point superamplitudes connected by an internal  (cut) propagator carrying momentum $l_1$. Treating $l_2'$ and $l_4$ as external momenta (they are on-shell due to the cut) we can follow the procedure of a four-point BCFW construction.  This involves rewriting fermionic  $\delta$-functions of both three-point amplitudes and integrating over $d^2 \eta_{l_1} d^2 \teta_{l_1}$, leading to the result 
\begin{alignat}{1} 
\label{quad-cut-as-double-l1-result}
\delta^4 (q_1 + q_{l_2'} + q_{l_4} + q_4 ) \delta^4 (\tq_{1 } + \tq_{l_2' } + \tq_{l_4 } + \tq_{4 } ) \;  w_{l_1}^a w_{l_1' a} \; \tw_{l_1}^{\da} \tw_{l_1' \da} \, .
\end{alignat}
Note that the $\delta$-functions are ensuring supermomentum conservation of the `external momenta' and that we do not have an internal propagator with momentum $l_1$ as in the recursive construction. Here, we get this propagator from uplifting the cut-expression for the one-loop amplitude. Furthermore, note that we do not have to shift any legs in order to use the BCFW prescription since the internal propagator is already on-shell due to the cut.

We may now perform the Grassmann integration over $\eta_{l_1}$ and $\teta_{l_1}$ in (\ref{quad-cut-as-double-l1-result}). Since the $w$-spinors are contracted we can simply use the spinor identity
\be
w_{l_1}^a w_{l_1' a} \tw_{l_1}^{\da} \tw_{l_1' \da} = - s_{l_4 l_2'}^{-1} = - s_{1 4}^{-1} 
\ , 
\ee
which is a direct generalisation of the corresponding result from the BCFW construction (see also appendix \ref{appB}).

We can now turn to the `upper' half of the cut-diagram. Following the description we derived above we get in a similar fashion after integrating over $\eta_{l_3}$ and $\teta_{l_3}$
\begin{alignat}{1}
\delta^4 (q_{l_2} + q_{2} + q_{3} + q_{l_4'} ) \delta^4 (\tq_{l_2 } + \tq_{2 } + \tq_{3 } + \tq_{l_4' } ) w_{l_3'}^a w_{l_3 a} \tw_{l_3'}^{\da} \tw_{l_3 \da}\,.
\end{alignat}
We also have 
\be
w_{l_1}^a w_{l_1' a} \tw_{l_1}^{\da} \tw_{l_1' \da} = - s_{l_2 l_4'}^{-1} = - s_{23}^{-1}\,.
\ee
Uplifting the quadruple cut, we get
\begin{alignat}{1}
A_{4;1} =  \int  \frac{d^6 l}{(2 \pi)^6}   & \int d^2 \eta_{l_2} d^2 \teta_{l_2} d^2 \eta_{l_4} d^2 \teta_{l_4} \bigg[  \frac{1}{l_1^2 l_2^2 l_3^2 l_4^2}   \frac{1}{ s_{1 4} s_{2 3}  }  \nonumber \\
& \hspace{0.4cm} \times  \delta^4 (q_1 + q_{l_2'} + q_{l_4} + q_4) \delta^4 (\tq_{1 } + \tq_{l_2' } + \tq_{l_4 } + \tq_{4 } )  \nonumber \\
& \hspace{0.4cm} \times \delta^4 (q_{l_2} + q_{2} + q_{3} + q_{l_4'} ) \delta^4 (\tq_{l_2 } + \tq_{2 } + \tq_{3 } + \tq_{l_4' } )  \bigg] \, .  
\end{alignat}
Since $l_i' = - l_i$ we can use the constraints given by the $\delta^4 (q_i)$ to eliminate the dependence of the remaining loop momenta in one of the sets of fermionic $\delta$-functions and write it as a sum over external momenta only. The same argument holds for the Grassmann functions $\delta^4 (\tq_{i})$, and we find 
\begin{alignat}{1}
A_{4;1} = \int  \frac{d^6 l}{(2 \pi)^6} &   \int d^2 \eta_{l_2} d^2 \teta_{l_2} d^2 \eta_{l_4} d^2 \teta_{l_4}   \bigg[   \delta^4 ( Q_{\mbox{\scriptsize ext}}) \delta^4 ( \tQ_{\mbox{\scriptsize ext}} )  \frac{1}{l_1^2 l_2^2 l_3^2 l_4^2}   \frac{1}{ s_{1 4} s_{2 4}  }\nonumber \\
& \hspace{0.5cm} \times \delta^4 (q_{l_2} + q_{2} + q_{3} - q_{l_4} ) \delta^4 (\tq_{l_2 } + \tq_{2 } + \tq_{3 } - \tq_{l_4 } )  \bigg],
\end{alignat}
where as before the $Q^A_{\mbox{\scriptsize ext}}$ and $\tQ_{A \;  \mbox{\scriptsize ext}}$ are the sums of all external supermomenta in $\eta$ and $\teta$ respectively. The remaining integrations over $\eta_{l_2}$ and $\eta_{l_4}$ and their $\teta$-counterparts yield just as in the case of the two-particle cut
\begin{alignat}{1}
& \epsilon_{A B C D} \; {l_2}^{A B} {l_4}^{C D} \;  \epsilon^{E F G H} p_{l_2 E F} p_{l_4 G H} =  64 \left( {l_2} \cdot {l_4} \right)^2.
\end{alignat}
The product of the two loop momenta cancels with the factor
\be
 s_{1 4} s_{2 3} = 2 (p_1 \cdot p_4) 2 (p_2 \cdot p_3) = 4 ({l_2'} \cdot {l_4})  ({l_2} \cdot {l_4'}) = (-1)^2 4 ({l_2} \cdot {l_4})^2 .
\ee 
so that our final result for the quadruple cut of the four-point superamplitude is 
\begin{alignat}{1}
A_{4;1} \propto \;  i st   A_{4;0} (1, \dots, 4) \int & \frac{d^6 l}{(2 \pi)^6} \bigg[  \frac{16}{ \;  l^2  (l+p_1)^2 (l + p_1 + p_2 )^2 (l-p_4)^2 } \bigg].
\end{alignat}
%%%%%%%%%%% -1 %%%%%%%%%%%%%%%%%%%%%
Hence we have shown that the quadruple cut gives the same structure as the two-particle cut discussed in Section \ref{sub:Two-particle-cuts}.

\subsubsection{Quadruple cut by Grassmann decomposition}

In this section we will calculate the quadruple cut of the one-loop four-point superamplitude in an alternate   fashion. Whereas in the last section we used the structure of the cut-expression to simplify the fermionic integrations, here we will explicitly perform the integrals by using constraints given by the $\delta$-functions.

To perform the Grassmann integrations we work directly at the level of the three-point superamplitudes. The quadruple cut results in the following four on-shell tree-level amplitudes (see Figure \ref{one-loop-quad-cut})
\be
A_{3}(l_{1},1, l'_{2}), \qquad  A_{3}( l_{2},2,l'_{3}), \qquad A_{3}(l_{3},3,l'_{4}), \qquad  A_{3}(l_{4},4,l'_{1}) \ .
\ee
Each of the three-point superamplitudes has the usual form \cite{siegel}
\be
A_{3,i}\  = \ i \left[\delta(Q_{i}^{A})\delta(\tilde{Q}_{iA})\right]^{2}\delta(W_{i})\delta(\tilde{W}_{i})
\ , 
\ee
where $i=1, \ldots,4 $ labels the corners. The arguments of the $\delta$-functions are 
\be
Q_{i}^{A}=q_{l_{i}}^{A}+q_{i}^{A}+q_{l'_{i+1}}^{A} \ , \qquad 
 W_{i}=w_{l_{i}}^{a}\eta_{l_{i}a}+w_{i}^{a}\eta_{ia}+w_{l'_{i+1}}^{a}\eta_{l'_{i+1}a}
 \ , 
\ee with the identification $l_{5}\equiv l_{1}$. Similar expressions hold for
for $\tilde{Q}_{iA}$ and $\tilde{W}_{i}$. Note that since $l_i' = - l_i$ we find it convenient to define spinors with primed momenta $l_i'$ as
\be
\lambda_{l_i'}^A = i \lambda_{l_i}^A \ , \qquad \tlambda_{l_i' A} = \tlambda_{l_i A} \ , \qquad    
\eta_{l_i '} = i \eta_{l_i} \ , \qquad 
 \teta_{l_i'} = i \teta_{l_i}
 \ , 
\ee
which we will frequently use in the following manipulations. 

We can use supermomentum conservation at each corner 
to reduce the number of $\delta$-functions depending on the loop variables $\eta_{l_i}$ and $\teta_{l_i}$. There is a choice involved and we choose to  remove the dependence of $\eta_{\ell_{i}}$ $(\tilde{\eta}_{\ell_{i}})$ from one copy of each $[\delta(Q_{i}^{A})\delta(\tilde{Q}_{iA})]^{2}$. This yields for the Grassmann integrations
\begin{alignat}{1} \label{4pt-quad-cut-Grassmann-int}
\delta^4 ( Q_{\mbox{\scriptsize ext }}) \delta^4 (\tQ_{\mbox{\scriptsize ext }}) \int \prod_{i=1}^{4} d^2 \eta_{l_i} d^2 \teta_{l_i} \bigg[ \delta (Q_i^A) \delta (\tQ_{i A}) \delta (W_i) \delta (\tW_i) \bigg] \ . 
\end{alignat}
We can simplify the calculation   by noticing that we have
to integrate over 16 powers of Grassmann variables (8 powers of $\eta$ and $\tilde{\eta}$ each) while at the same time we have 16 $\delta$-functions in total. Therefore, when expanding the fermionic functions, each of them must contribute a power of Grassmann variables we are going to integrate over. Unless this is so,
the result is zero. In other words, we can only pick the terms in
the $\delta$-functions that contribute an $\eta_{l_{i}}$ or $\tilde{\eta}_{l_{i}}$.
This simplifies the structure considerably  as we can drop all terms depending on external variables.  

Equation (\ref{4pt-quad-cut-Grassmann-int}) now becomes
\begin{alignat}{1} \label{4pt-quad-cut-Grassmann-final}
\delta^4 ( Q_{\mbox{\scriptsize ext }}) \delta^4 (\tQ_{\mbox{\scriptsize ext }}) \int \prod_{i=1}^{4} d^2 \eta_{l_i} d^2 \teta_{l_i} \bigg[ & \delta (q_{l_i}^A - q_{l_{i+1}}^A) \;\delta (\tq_{l_i A} - \tq_{l_{i+1} A} )  \\
& \hspace{-0.5cm} \times  \delta (w_{l_i}^a \eta_{l_i a} + i w_{l_{i+1}'}^a \eta_{l_{i+1} a} ) \delta (\tw_{l_i}^{\da} \teta_{l_i \da} + i \tw_{l_{i+1}'}^{\da } \teta_{l_{i+1} \da} ) \bigg] \ . \nonumber 
\end{alignat}
Notice that the $w$-spinors $w_{l_{i+1}'}^a$ are not identical to $w_{l_{i+1}}^a$. 

Since the $\delta$-functions only depend on the $\eta_{l_i}$ and $\teta_{l_i}$, we find convenient to decompose the integration variables as
\be
\eta_{l_i}^a = u_{l_i}^a \eta_{l_i}^{\parallel} + w_{l_i}^a \eta_{l_i}^{\perp} \, , \qquad 
 \teta_{l_i}^{\da} = \tu_{l_i}^{\da} \teta_{l_i}^{\parallel} + \tw_{l_i}^{\da} \teta_{l_i}^{\perp} 
 \ , 
\ee
which implies 
\be
w_{l_i a} \eta_{l_i}^a = \eta_{l_i}^{\parallel} \, , \qquad  u_{l_i}^a \eta_{l_i a} = \eta_{l_i}^{\perp} \ .  
\ee
Hence, we can rewrite the arguments of the $\delta$-functions in the $w$-spinors as
\begin{alignat}{1}
i w_{l_{i+1}'}^a \eta_{l_{i+1} a} & = i w_{l_{i+1}'}^a \left( u_{l_{i+1} a} \eta_{l_{i+1} }^{\parallel} + w_{l_{i+1} a} \eta_{l_{i+1}}^{\perp} \right) = \frac{i}{\sqrt{-s_{i, i+1}}} u_{l_{i+1}}^a w_{l_{i+1} a} \eta_{l_{i+1}}^{\perp} \nonumber \\
&  = \frac{i}{\sqrt{-s_{i, i+1}}} \eta_{l_{i+1}}^{\perp} 
\ , 
\end{alignat}
and similarly we have $i \tw_{l_{i+1}'}^{\da} \teta_{l_{i+1} \da} = \frac{i}{\sqrt{-s_{i, i+1}}} \teta_{l_{i+1}}^{\perp}$. 
Notice that we have used  the fact that the $w_{l_{i+1}}'$ can be normalised such that they are proportional to the $u_{l_{i+1}}$ if the momenta fulfill the condition $l_{i+1}' = - l_{i+1}$. We give some more detail on such relations in  Appendix \ref{appB2}. 

Using this, the $\delta$-functions in the $w$-spinors become
\be
\delta \big( - \eta_{l_i}^{\parallel} + \frac{i}{\sqrt{-s_{i, i+1}}} \eta_{l_{i+1}}^{\perp}   \big) \delta \big( - \teta_{l_i}^{\parallel} + \frac{i}{\sqrt{-s_{i, i+1}}} \teta_{l_{i+1}}^{\perp}   \big) \ .
\ee
Next we proceed by    integrating first  over the $\eta_{l_i}^{\parallel}$ variables. This sets 
\be \label{solution_eta_parallel}
\eta_{l_i}^{\parallel} =  \frac{-i}{\sqrt{-s_{i, i+1}}} \eta_{l_{i+1}}^{\perp}\, ,
\ee
with similar expressions for the $\teta_{l_i}^{\parallel}$. 
We then plug this into the remaining $\delta$-functions of  \eqref{4pt-quad-cut-Grassmann-final}. 
First we notice that
\begin{alignat}{1}
& \delta \big( \lambda_{l_{i}}^{Aa} \eta_{l_{i} a } + \lambda_{l'_{i+1}}^{Aa} \eta_{l'_{i+1}a} \big) \delta \big( \tlambda_{l_{i}}^{A \da} \teta_{l_{i} \da} + \tlambda_{l'_{i+1}}^{A \da} \teta_{l'_{i+1} \da} \big)   \nonumber\\
 =&  \la l_{i}^a | {l'}_{i+1}^{\da} ] \eta_{l_{i} a } \teta_{l'_{i+1} \da} + \la {l'}_{i+1}^a | l_i^{\da} ] \eta_{l'_{i+1} a} \teta_{ l_{i} \da} = - u_{l_i}^a \tu_{l_{i+1}'}^{\da} \eta_{l_{i} a } \teta_{l'_{i+1} \da} + u_{l_{i+1}'}^{a} \tu_{l_i}^{\da} \eta_{l'_{i+1} a} \teta_{ l_{i} \da}  \ .
\end{alignat}
The decomposition of the Grassmann spinors then yields
\be
u_{l'_{i+1}}^a \eta_{l'_{i+1} a} = i \sqrt{-s_{i, i+1}} w_{l_{i+1}}^a u_{l_{i+1} a} \eta_{l_{i+1}}^{\parallel} = -i \sqrt{-s_{i, i+1}} \; \eta_{l_{i+1}}^{\parallel}
\ee
and
\be
\tu_{l'_{i+1}}^{\da} \teta_{l'_{i+1} \da} = - i \sqrt{-s_{i, i+1}} \; \teta_{l_{i+1}}^{\parallel} \ .
\ee
The remaining Grassmann integrations give
\begin{alignat}{1}
& \int \prod_{i=1}^4 d \eta_{l_i}^{\perp} d \teta_{l_i}^{\perp} \bigg[ \delta \big( \lambda_{l_{i}}^{Aa} \eta_{l_{i} a } + \lambda_{l'_{i+1}}^{Aa} \eta_{l'_{i+1}a} \big) \delta \big( \tlambda_{l_{i}}^{A \da} \teta_{l_{i} \da} + \tlambda_{l'_{i+1}}^{A \da} \teta_{l'_{i+1} \da} \big) \bigg]   \nonumber \\
= &  \int \prod_{i=1}^4 d \eta_{l_i}^{\perp} d \teta_{l_i}^{\perp} \bigg[ i \sqrt{-s_{i, i+1}} \; \eta_{l_i}^{\perp}  \teta_{l_{i+1}}^{\parallel} - i \sqrt{-s_{i, i+1}} \; \eta_{l_{i+1}}^{\parallel}  \teta_{l_{i}}^{\perp} \bigg] \nonumber \\
= & \int \prod_{i=1}^4 d \eta_{l_i}^{\perp} d \teta_{l_i}^{\perp} \bigg[  \eta_{l_i}^{\perp} \teta_{l_{i+2}}^{\perp} - \eta_{l_{i+2}}^{\perp }  \teta_{l_{i}}^{\perp} \bigg]
\ , 
\end{alignat}
where we have used the solutions for $\eta_{l_i}^{\parallel}$ and $\teta_{l_i}^{\parallel}$ following (\ref{solution_eta_parallel}). The integration is now straightforward, since the integrand is simply  given by
\begin{alignat}{1}
&  \big( \eta_{l_1}^{\perp} \teta_{l_3}^{\perp} - \eta_{l_3}^{\perp} \teta_{l_1}^{\perp} \big) \big( \eta_{l_2}^{\perp} \teta_{l_4}^{\perp} - \eta_{l_4}^{\perp} \teta_{l_2}^{\perp} \big) \big( \eta_{l_3}^{\perp} \teta_{l_1}^{\perp} - \eta_{l_1}^{\perp} \teta_{l_3}^{\perp} \big) \big( \eta_{l_4}^{\perp} \teta_{l_2}^{\perp} - \eta_{l_2}^{\perp} \teta_{l_4}^{\perp} \big) \nonumber \\
 = & \;  4 \eta_{l_1}^{\perp} \teta_{l_3}^{\perp} \eta_{l_2}^{\perp} \teta_{l_4}^{\perp} \eta_{l_3}^{\perp} \teta_{l_1}^{\perp} \eta_{l_4}^{\perp} \teta_{l_2}^{\perp} \ . 
\end{alignat}
This yields 
\begin{alignat}{1}
A_{4;1} \propto - 4 i s t A_{4;0} (1, \dots, 4) \int & \frac{d^6 l}{(2 \pi)^6} \bigg[  \frac{1}{ \;  l^2  (l+p_1)^2 (l + p_1 + p_2 )^2 (l-p_4)^2 } \bigg]
\ , 
\end{alignat}
recovering the expected result of \cite{siegel} from two-particle cuts.

\section{One-loop five-point superamplitude} 
\label{5555}

\begin{figure}
\begin{center}
\includegraphics[width=8cm]{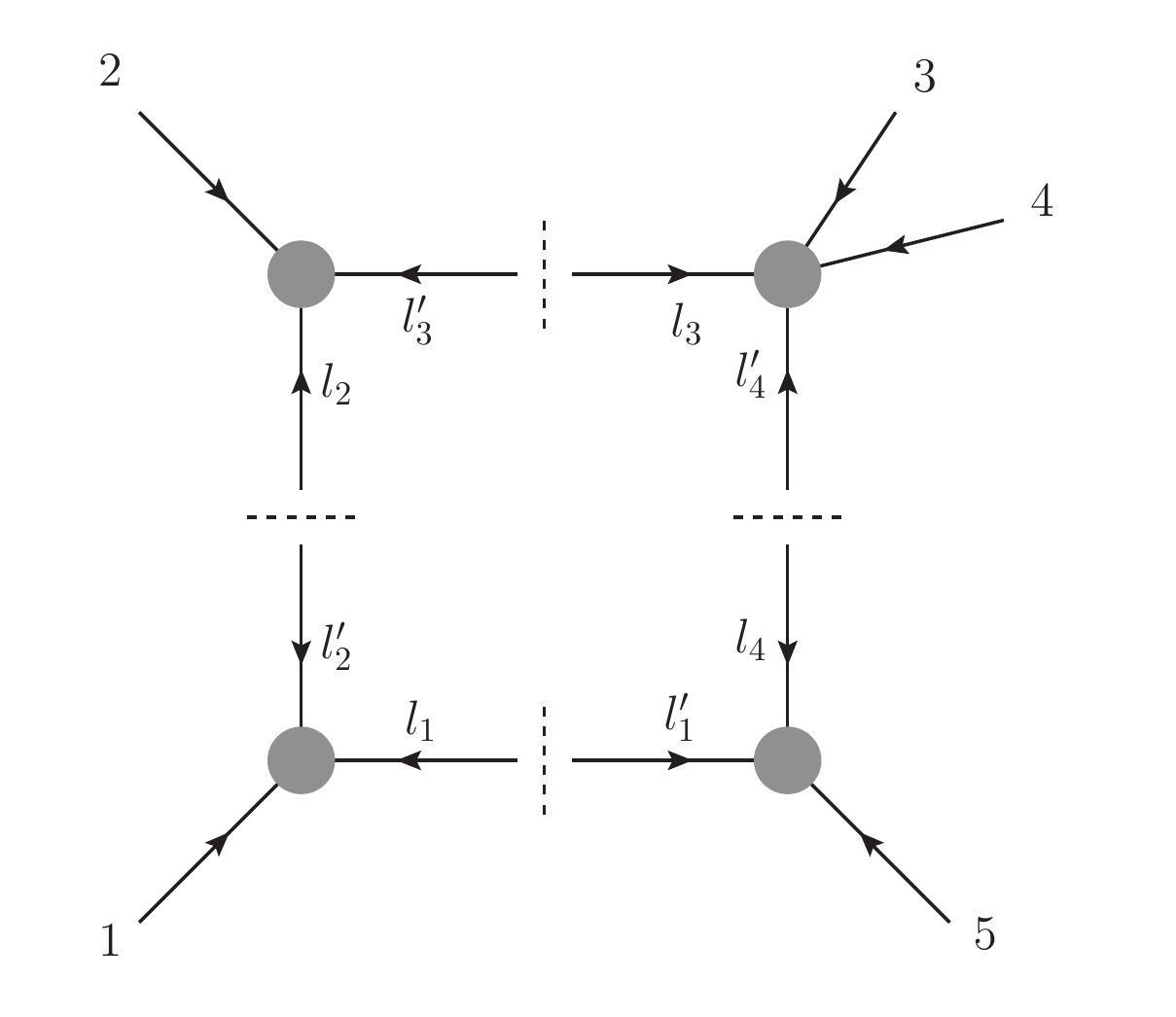}
\caption{\it A specific quadruple cut of a five-point superamplitude. We choose to cut the legs such that we have the massive corner for momenta $p_3, p_4$.}
\label{5pt-one-loop-quad-cut}
\end{center}
\end{figure}

We now move on to the one-loop five-point superamplitude and calculate its quadruple cuts. 
These cuts will reveal the presence of a  linear pentagon integral, which we will reduce using standard Passarino-Veltman (PV) techniques to  a scalar pentagon plus  scalar box integrals.  Note that  we are considering here one-loop amplitudes in
the maximally supersymmetric theory in six dimensions which are free of IR and UV divergences. Therefore,
bubbles and triangles which would be UV divergent in six dimensions must be absent.

\subsection{Quadruple cuts}

The quadruple cut we consider has the structure
\beqa
\hspace{-1.5cm}A_{5;1} \big|_{\mbox{\scriptsize (3,4)-cut}} &= & 
\int \frac{d^6 l}{(2 \pi)^6} \delta^+ (l_1^2) \, \delta^+ (l_2^2)\, \delta^+ (l_3^2) \, \delta^+ (l_4^2)
\\
&&
A_3 (l_1, p_1, -l_2) \,   A_3 (l_2, p_2, -l_3) \, 
   A_4 (l_3, p_3, p_4, -l_4) \,   A_3 (l_4, p_5 , -l_1)  \, , 
\nonumber 
\eeqa
where the subscript $(3,4)$ indicates where the massive corner is located, see Figure \ref{5pt-one-loop-quad-cut}.  In the following we will discuss this specific cut and all other cuts can be treated in an identical way.

From the three three-point superamplitudes and the four-point superamplitude, we have the following fermionic $\delta$-functions, 
\begin{alignat}{1}
 \left[ \delta (Q_1^A)  \delta (\tQ_{1 A}) \right]^2  \delta (W_1) \delta ( \tW_1 ) & \left[ \delta (Q_2^B)  \delta (\tQ_{2 B}) \right]^2  \delta (W_2) \delta ( \tW_2 )  \nonumber \\
\times \;   \delta^4 (Q_3^C) \delta^4 (\tQ_{3 C} ) &  \left[ \delta (Q_4^D)  \delta (\tQ_{4 D}) \right]^2 \delta (W_4) \delta ( \tW_4 )
\ , 
\end{alignat}
where the $Q_i^A$ and the $W_i$ are defined as sums over the supermomenta and products of $w$- and $\eta$-spinors respectively at a given corner (including internal legs). We may now use the supermomentum constraints $Q_i^A  = 0$ at all four corners and rewrite the $\delta^4 (Q_3) \delta^4 (\tQ_{3 } )$ as a total $\delta^8$ in the external momenta only, 
\be
\delta^4 (Q_3) =  \delta^4 (Q_3 + Q_1 + Q_2+ Q_4 ) = \delta^4 ( Q_{\mbox{\scriptsize ext}}).
\ee
One is then left with the Grassmann integrations 
\begin{alignat}{1}
\int \prod_{i=1}^4 d^2 \eta_{l_i} d^2 \teta_{l_i} & \bigg\{ \left[ \delta ( q_{l_1}^A + q_1^A - q_{l_2}^A )  \delta ( \tq_{l_1 A} + \tq_{1 A} - \tq_{l_2 A} ) \right]^2 \nonumber \\
& \;  \delta (w_{l_1}^a \eta_{l_1 a} + w_1^a \eta_{1 a} + i w_{l_2'}^a \eta_{l_2 a})  \delta (\tw_{l_1}^{\da} \teta_{l_1 \da} + \tw_1^{\da} \teta_{1 \da} + i \tw_{l_2'}^{\da} \teta_{l_2 \da}) \nonumber \\
& \left[ \delta ( q_{l_2}^B + q_2^B - q_{l_3}^B )  \delta ( \tq_{l_2 B} + \tq_{2 B} - \tq_{l_3 B} ) \right]^2 \nonumber \\
& \;  \delta (w_{l_2}^b \eta_{l_2 b} + w_2^b \eta_{2 b} + i w_{l_3'}^b \eta_{l_3 b})  \delta (\tw_{l_2}^{\db} \teta_{l_2 \db} + \tw_2^{\db} \teta_{2 \db} + i \tw_{l_3'}^{\db} \teta_{l_3 \db}) \nonumber \\
& \left[ \delta ( q_{l_4}^D + q_5^D - q_{l_1}^D )  \delta ( \tq_{l_4 D} + \tq_{5 D} - \tq_{l_1 D} ) \right]^2 \nonumber \\
& \;  \delta (w_{l_4}^c \eta_{l_4 c} + w_5^c \eta_{5 c} + i w_{l_1'}^c \eta_{l_1 c})  \delta (\tw_{l_4}^{\dc} \teta_{l_4 \dc} + \tw_5^{\dc} \teta_{5 \dc} + i \tw_{l_1'}^{\dc} \teta_{l_1 \dc}) \bigg\} \ . 
\end{alignat}
Unfortunately, a decomposition as used for the quadruple cut of the four-point one-loop superamplitude is not immediately useful here. 
%One notes that we have more powers of Grassmann spinors in the integrand than one has to integrate over. Hence, a %decomposition would not lead to any simplifications and we have to expand the fermionic functions. 
However, we notice that, due to the particular dependence of the $\delta$-functions on the loop momenta $l_i$,    by removing a total $\delta^8$ from the integrand one can restrict  the dependence on the Grassmann variables $\eta_{l_3}$ and $\eta_{l_4}$ to six  $\delta$-functions each for this specific cut. This allows us to narrow the possible combinations of coefficients for, say,  two powers of $\eta_{l_4 a}$ and two powers of $\teta_{l_4 \da}$. For example, two powers of $\eta_{l_4 a}$ can either come both from $\delta (Q_4^A) \delta (Q_4^B) $ or one from $\delta (Q_4^A)$ and one from%
\footnote{ This is similar to the recursive calculation of the five-point tree-level superamplitude in six dimensions, see also \cite{siegel}.} 
$\delta (W_4)$,  and both possibilities needs to be appropriately  contracted with the possible combinations from  $\delta (\tQ_{4 A}) \delta ( \tQ_{4 B}) \delta (\tW_4)$.
If we choose both powers of $\eta_{l_4 a}$  from $\delta (Q_4^A) \delta (Q_4^B) $  we have a coefficient
\be
\lambda_{l_4}^{A a} \eta_{l_4 a} \lambda_{l_4}^{B b} \eta_{l_4 b}
\ , 
\ee
which  will be contracted at least by a $\tlambda_{l_4 A}^{\da}$ or $\tlambda_{l_4 B}^{\da}$ coming from the possible combinations for $\teta_{l_4 \da}$. Since $\lambda^A_{i a} \tlambda_{i A \da} = 0$ these terms vanish.  

In conclusion, the only non-vanishing combination is
\be
\lambda_{l_4}^{A a} \eta_{l_4 a} \delta (\tq_{5 A} - \tq_{l_1 A}) \delta (q_5^B - q_{l_1}^B) \tlambda_{l_4 B}^{\da} \teta_{l_4 \da} w_{l_4}^b \eta_{l_4 b} \tw_{l_4}^{\db} \teta_{l_4 \db}
\ . 
\ee
%%%%%%%%%%%% 2 %%%%%%%%%%%%%%%%
The same argument holds for the expansion of the $\delta$-functions depending on $\eta_{l_3 a}$ and $\teta_{l_3 \da}$. Here, we only have to deal with additional signs and factors of $i$. We get for the expansion 
\be
(-1) \lambda_{l_3}^{A a} \eta_{l_3 a} \delta (\tq_{l_2 A} + \tq_{2 A}) \delta (q_{l_2}^B + q_{2}^B) (-1) \tlambda_{l_3 B}^{\da} \teta_{l_3 \da} i w_{l_3'}^b \eta_{l_3 b} i \tw_{l_3'}^{\db} \teta_{l_3 \db} \, . 
\ee
This leads us to the structure
\begin{alignat}{1}
\int \prod_{i=1}^4 d^2 \eta_{l_i} d^2 \teta_{l_i} \bigg\{ & \left[ \delta ( q_{l_1}^A + q_1^A - q_{l_2}^A )  \delta ( \tq_{l_1 A} + \tq_{1 A} - \tq_{l_2 A} ) \right]^2 \nonumber \\
& \;  \delta (w_{l_1}^a \eta_{l_1 a} + w_1^a \eta_{1 a} + i w_{l_2'}^a \eta_{l_2 a})  \delta (\tw_{l_1}^{\da} \teta_{l_1 \da} + \tw_1^{\da} \teta_{1 \da} + i \tw_{l_2'}^{\da} \teta_{l_2 \da}) \nonumber \\
& \lambda_{l_3}^{C c} \eta_{l_3 c} \delta (\tq_{l_2 C} + \tq_{2 C}) \delta (q_{l_2}^D + q_{2}^D)  \tlambda_{l_3 D}^{\dc} \teta_{l_3 \dc} (i)^2 w_{l_3'}^c \eta_{l_3 c}  \tw_{l_3'}^{\dc} \teta_{l_3 \dc} \nonumber \\
& \lambda_{l_4}^{E d} \eta_{l_4 d} \delta (\tq_{5 E} - \tq_{l_1 E}) \delta (q_5^F - q_{l_1}^F) \tlambda_{l_4 F}^{\dd} \teta_{l_4 \dd} w_{l_4}^d \eta_{l_4 d} \tw_{l_4}^{\dd} \teta_{l_4 \dd} \bigg\} \, .
\end{alignat}
Notice that we have not expanded the six $\delta$-functions of the first corner yet, therefore  we still have supermomentum conservation $Q^A_1 = 0$, $\tQ_1^A = 0$. We can use this constraint to remove the dependence on $\eta_{l_2 a}$ in the third line of the above integrand, using
$q_{l_2}^A = q_{l_1}^A + q_1^A$.
Our fermionic integral then becomes
\begin{alignat}{1}
\int \prod_{i = 1}^4 d^2 \eta_{l_i} d^2 \teta_{l_i} \bigg\{ & \left[ \delta ( q_{l_1}^A + q_1^A - q_{l_2}^A )  \delta ( \tq_{l_1 A} + \tq_{1 A} - \tq_{l_2 A} ) \right]^2 \nonumber \\
& \;  \delta (w_{l_1}^a \eta_{l_1 a} + w_1^a \eta_{1 a} + i w_{l_2'}^a \eta_{l_2 a})  \delta (\tw_{l_1}^{\da} \teta_{l_1 \da} + \tw_1^{\da} \teta_{1 \da} + i \tw_{l_2'}^{\da} \teta_{l_2 \da}) \nonumber \\
& (i)^2  \eta_{l_3 c}  \teta_{l_3 \dc}  \eta_{l_3 c'}  \teta_{l_3 \dc'}    \lambda_{l_3}^{C c} \tlambda_{l_3 D}^{\dc} w_{l_3'}^{c'}  \tw_{l_3'}^{\dc'} \delta (\tq_{l_1 C} + \tq_{1 C}+ \tq_{2 C}) \delta (q_{l_1}^D + q_{1}^D + q_{2}^D)   \nonumber \\
&  \eta_{l_4 d} \teta_{l_4 \dd} \eta_{l_4 d'} \teta_{l_4 \dd'} \lambda_{l_4}^{E d} \tlambda_{l_4 F}^{\dd}  w_{l_4}^{d'}    \tw_{l_4}^{\dd'} \delta (\tq_{5 E} - \tq_{l_1 E}) \delta (q_5^F - q_{l_1}^F) \bigg\} \, .
\end{alignat}
We immediately see that, just as before, only the first six $\delta$-functions depend on $\eta_{l_2 a} $ and $ \teta_{l_2 \da}$ so we can expand straight away (noticing that we get another factor of $(i)^2$ from this expansion)
\begin{alignat}{1}
\int \prod_{i = 1}^4 d^2 \eta_{l_i} d^2 \teta_{l_i} \bigg\{ &  (i)^2  \eta_{l_2 b}  \teta_{l_2 \db}  \eta_{l_2 b'}  \teta_{l_2 \db'}    \lambda_{l_2}^{A b} \tlambda_{l_2 B}^{\db} w_{l_2'}^{b'}  \tw_{l_2'}^{\db'}  \delta (\tq_{l_1 A} + \tq_{1 A} ) \delta (q_{l_1}^B + q_1^B)  \nonumber \\
& (i)^2  \eta_{l_3 c}  \teta_{l_3 \dc}  \eta_{l_3 c'}  \teta_{l_3 \dc'}    \lambda_{l_3}^{C c} \tlambda_{l_3 D}^{\dc} w_{l_3'}^{c'}  \tw_{l_3'}^{\dc'} \delta (\tq_{l_1 C} + \tq_{1 C}+ \tq_{2 C}) \delta (q_{l_1}^D + q_{1}^D + q_{2}^D)   \nonumber \\
&  \eta_{l_4 d} \teta_{l_4 \dd} \eta_{l_4 d'} \teta_{l_4 \dd'} \lambda_{l_4}^{E d} \tlambda_{l_4 F}^{\dd}  w_{l_4}^{d'}    \tw_{l_4}^{\dd'} \delta (\tq_{5 E} - \tq_{l_1 E}) \delta (q_5^F - q_{l_1}^F) \bigg\} \, .
\end{alignat}
One notes that, by expanding the fermionic $\delta$-functions, the dependence on the Grassmann parameters $\eta_{l_1 a}$ and $\teta_{l_1 \da}$ has reduced to 
\be \label{etal1-delta-structure}
\delta (\tq_{l_1 A} + \tq_{1 A} ) \delta (q_{l_1}^B + q_1^B) \delta (\tq_{l_1 C} + \tq_{1 C}+ \tq_{2 C}) \delta (q_{l_1}^D + q_{1}^D + q_{2}^D) \delta (\tq_{5 E} - \tq_{l_1 E}) \delta (q_5^F - q_{l_1}^F)
\ee
only.
 Expanding this further gives the sought-after coefficient of $\eta_{l_1 a} \eta_{l_1 b} \teta_{l_1 \da} \teta_{l_1 \db} $. The result (in an appropriate order of the Grassmann spinors) of the expansion of the six $\delta$-functions in (\ref{etal1-delta-structure}) is then given by
\begin{alignat}{1}
& \teta_{l_1 \da} \teta_{l_1 \db}  \bigg(  - \teta_{1 \dc}   \tlambda_{l_1 A}^{\da} \tlambda_{1 C}^{\dc}  \tlambda_{l_1 E}^{\db}  - \teta_{2 \dc}   \tlambda_{l_1 A}^{\da} \tlambda_{2 C}^{\dc} \tlambda_{l_1 E}^{\db}  - \teta_{5 \dc}    \tlambda_{l_1 A}^{\da} \tlambda_{l_1 C}^{\db} \tlambda_{5 E}^{\dc}  +  \teta_{1 \dc}    \tlambda_{1 A}^{\dc} \tlambda_{l_1 C}^{\da}  \tlambda_{l_1 E}^{\db} \bigg) \nonumber \\
 \times \;  &  \eta_{l_1 a} \eta_{l_1 b} \bigg( \eta_{5 c} \lambda_{l_1}^{B a} \lambda_{l_1}^{D b} \lambda_{5}^{F c} +  \eta_{1 c} \lambda_{l_1}^{B a} \lambda_{1}^{D c} \lambda_{l_1}^{F b} +  \eta_{2 c} \lambda_{l_1}^{B a} \lambda_{2}^{D c} \lambda_{l_1}^{F b} -  \eta_{1 c} \lambda_{1}^{B c} \lambda_{l_1}^{D a} \lambda_{l_1}^{F b}   \bigg) \, .
\end{alignat}
Having extracted the right powers of the Grassmann variables from all fermionic $\delta$-functions, we can now integrate over the $\eta_{l_i}$ and $\teta_{l_i}$. The integration is straightforward and yields, 
\begin{alignat}{1} \label{5pt-quad-result-grassmann-int}
& \bigg( \;   \teta_{1 \dc} [1^{\dc} | l_3 \ra \cdot w_{l_3'}\;  w_{l_2'} \cdot \la l_2 | \hal_1 | l_4 \ra \cdot w_{l_4} - \teta_{1 \dc} [1^{\dc} | l_2 \ra \cdot w_{l_2'}\;  w_{l_3'} \cdot \la l_3 | \hal_1 | l_4 \ra \cdot w_{l_4} \nonumber \\
& + \;    \teta_{2 \dc} [2^{\dc} | l_3 \ra \cdot w_{l_3'}\;  w_{l_2'} \cdot \la l_2 | \hal_1 | l_4 \ra \cdot w_{l_4}  + \teta_{5 \dc} [5^{\dc} | l_4 \ra \cdot w_{l_4}\;  w_{l_2'} \cdot \la l_2 | \hal_1 | l_3 \ra \cdot w_{l_3'} \bigg) \nonumber \\
\times &  \bigg( \;  \eta_{1 c} \la 1^c |Êl_3 ] \cdot \tw_{l_3'} \;  \tw_{l_2'} \cdot [ l_2 | \hal_1 | l_4 ] \cdot \tw_{l_4} -  \eta_{1 c} \la 1^c |Êl_2 ] \cdot \tw_{l_2'} \;  \tw_{l_3'} \cdot [ l_3 | \hal_1 | l_4 ] \cdot \tw_{l_4} \nonumber \\
& + \;  \eta_{2 c} \la 2^c |Êl_3 ] \cdot \tw_{l_3'} \;  \tw_{l_2'} \cdot [ l_2 | \hal_1 | l_4 ] \cdot \tw_{l_4} +  \eta_{5 c} \la 5^c |Êl_4 ] \cdot \tw_{l_4} \;  \tw_{l_2'} \cdot [ l_2 | \hal_1 | l_3 ] \cdot \tw_{l_3'} \bigg) \, . 
\end{alignat}
Here we introduced the notation that $ w_{l_i} \cdot \la l_i | := w_{l_i}^a \la l_{i, a} | $, and the $\hal_i$ are slashed momenta, with e.g.
\be
w_{l_2'} \cdot \la l_2 | \hal_1 | l_3 \ra \cdot w_{l_3'} = w_{l_2'}^a \lambda_{l_2, a}^A l_{1, A B} \lambda_{l_3, b}^B w_{l_3'}^b \, .
\ee
Next, one rewrites the spinor expressions in (\ref{5pt-quad-result-grassmann-int}) in terms of  six-dimensional momenta, thereby  removing any dependence on $u$- and $w$-spinors. An important observation to do so is the fact that the expressions depending on $\teta_1$ and/or $\eta_1$ antisymmetrise among themselves\footnote{We give more details on these manipulations in Appendix \ref{appB2}.}. The result of these manipulations is 
\begin{alignat}{1} 
\label{5pt-quad-result-prePV}
& \teta_{1 \dc} \eta_{1 c} \frac{1}{s_{12}} [ 1^{\dc} | \hap_2 \hal_1 \hap_5  \hap_2 |Ê1^c \ra
 - \teta_{1 \dc} \eta_{2 c} \frac{1}{s_{12}} [ 1^{\dc} | \hap_2 \hap_5 \hal_1 \hap_1 |Ê2^c \ra + \teta_{2 \dc} \eta_{1 c} \frac{1}{s_{12}} [ 2^{\dc} | \hap_1 \hal_1 \hap_5  \hap_2 |Ê1^c \ra \nonumber \\
+ &  \teta_{1 \dc} \eta_{5 c}  [ 1^{\dc} | \hap_2 \hal_1 |Ê5^c \ra - \teta_{5 \dc} \eta_{1 c}  [ 5^{\dc} | \hal_1 \hap_2  |Ê1^c \ra + \teta_{2 \dc} \eta_{2 c} \frac{1}{s_{12}} [ 2^{\dc} | \hap_1 \hal_1 \hap_5 \hap_1 |Ê2^c \ra   \nonumber \\
+ & \teta_{5 \dc} \eta_{5 c} \frac{1}{s_{15}} [ 5^{\dc} | \hap_1 \hal_1 \hap_2 \hap_1 | 5^c \ra + \teta_{5 \dc} \eta_{2 c}  [ 5^{\dc} | \hal_1  \hap_1 |Ê2^c \ra - \teta_{2 \dc} \eta_{5 c}  [ 2^{\dc} | \hap_1 \hal_1 |Ê5^c \ra \, . 
\end{alignat}
\subsection{Final result (before PV reduction)}
Including all appropriate prefactors, our result for the five-point one-loop superamplitude is expressed in terms of a single integral function, namely a linear pentagon integral. Explicitly, 
\begin{alignat}{1} \label{result-5pt-1loop}
A_{5;1}\  = \ \mathcal{C}_{\mu}  \, I^{\mu}_{5, l_1}\ , 
\end{alignat}
where 
\be
\label{lin_pentagon}
I^{\mu}_{5, l_1} (1, \dots, 5) := \int \frac{d^D l}{(2 \pi)^D}  \frac{l_1^{\mu}}{ l_1^2 l_2^2 l_3^2 (p_3 +l_3 )^2 l_5^2} \, , 
\ee
is the linear pentagon, and  the coefficient $\mathcal{C}_{\mu}$ is given by
\begin{alignat}{1} 
\mathcal{C}_{\mu} = & \frac{1}{s_{34}} \bigg\{ \teta_{1 \dc} \eta_{1 c} \frac{1}{s_{12}} [ 1^{\dc} | \hap_2 \hat{\sigma}_{\mu} \hap_5  \hap_2 |Ê1^c \ra
 - \teta_{1 \dc} \eta_{2 c} \frac{1}{s_{12}} [ 1^{\dc} | \hap_2 \hap_5 \hat{\sigma}_{\mu} \hap_1 |Ê2^c \ra + \teta_{2 \dc} \eta_{1 c} \frac{1}{s_{12}} [ 2^{\dc} | \hap_1 \hat{\sigma}_{\mu} \hap_5  \hap_2 |Ê1^c \ra \nonumber \\
+ & \;  \teta_{1 \dc} \eta_{5 c}  [ 1^{\dc} | \hap_2 \hat{\sigma}_{\mu} |Ê5^c \ra - \teta_{5 \dc} \eta_{1 c}  [ 5^{\dc} | \hat{\sigma}_{\mu} \hap_2  |Ê1^c \ra + \teta_{2 \dc} \eta_{2 c} \frac{1}{s_{12}} [ 2^{\dc} | \hap_1 \hat{\sigma}_{\mu}\hap_5 \hap_1 |Ê2^c \ra   \nonumber \\
+ & \;  \teta_{5 \dc} \eta_{5 c} \frac{1}{s_{15}} [ 5^{\dc} | \hap_1 \hat{\sigma}_{\mu} \hap_2 \hap_1 | 5^c \ra + \teta_{5 \dc} \eta_{2 c}  [ 5^{\dc} | \hat{\sigma}_{\mu} \hap_1 |Ê2^c \ra - \teta_{2 \dc} \eta_{5 c}  [ 2^{\dc} | \hat{\sigma}_{\mu} \hal_1 |Ê5^c \ra \bigg\}Ê\ . 
\end{alignat}
The factor of $1/{s_{34}}$ and the additional propagator in the pentagon appearing in \eqref{result-5pt-1loop}, are due to the prefactor of the four-point tree-level superamplitude entering the cut. We now proceed and summarise the  result of the PV reduction of  \eqref{lin_pentagon} in the next section. 

\begin{figure}[t]
\begin{center}
\includegraphics[width=8cm]{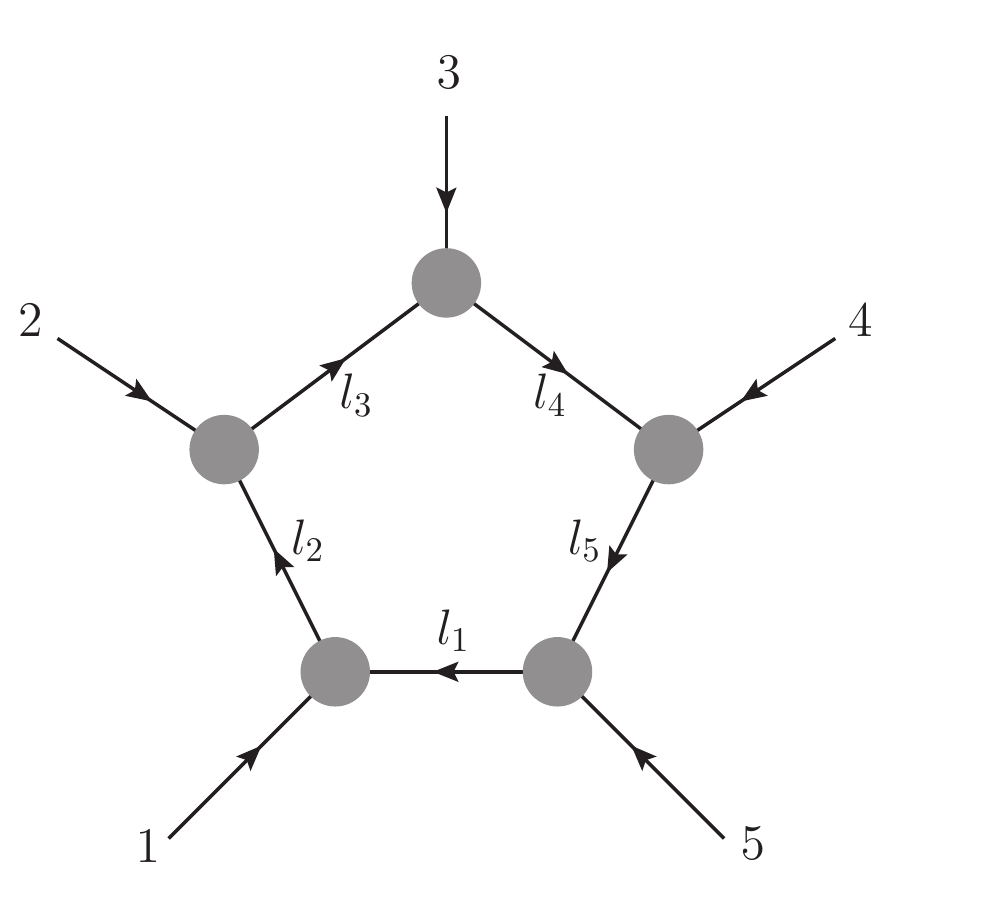}
\caption{\it A generic pentagon loop integral.}
\label{generic-pentagon}
\end{center}
\end{figure}

\subsection{Final result (after PV reduction)}
\label{pvred}

The PV reduction of  \eqref{lin_pentagon}  allows us to re-express a linear pentagon in terms of a scalar pentagon and  scalar box functions.
Using this, we re-express  the one-loop five-point superamplitude  as
\begin{alignat}{1} \label{5pt-one-loop-expansion}
A_{5;1} = \mathcal{C}^{(5)} I_5 (1, \dots, 5) + \sum_{i=1}^5 \mathcal{C}^{(4,i)} I_{4,i} (1, \dots, 5)
\ , 
\end{alignat}
where we introduced the scalar integral functions $I_5$ for the pentagon and $I_{4,i}$ for the boxes. Here, the index $i$ in  $I_{4,i}$ labels the first leg of the massive corner for a clockwise ordering of the external states.

Explicitly, the coefficients for the specific cut we discussed in the previous section are given by

\begin{alignat}{1}
\mathcal{C}^{(5) / (4,3)}  = & \;   \teta_{1 \dc} \eta_{1 c} \frac{1}{s_{12} } \big( s_{12} [ 1^{\dc} | \hap_5 \hap_2 |Ê1^c \ra A^{(5) / (4,3)} + [ 1^{\dc} | \hap_2 \hap_3  \hap_5 \hap_2 |Ê1^c \ra C^{(5) / (4,3)} \big) \nonumber \\
+ & \; \teta_{2 \dc} \eta_{2 c} \frac{1}{s_{12}} \big( s_{12} [ 2^{\dc} | \hap_5 \hap_1 |Ê2^c \ra B^{(5) / (4,3)} + [ 2^{\dc} | \hap_1 \hap_3 \hap_5 \hap_1 |Ê2^c \ra C^{(5) / (4,3)} \big)  \nonumber \\
+ & \; \teta_{5 \dc} \eta_{5 c} \frac{1}{s_{15}} \big(  [ 5^{\dc} | \hap_1 \hap_3 \hap_2 \hap_1 | 5^c \ra C^{(5) / (4,3)} +  s_{15} [ 5^{\dc} | \hap_2 \hap_1 | 5^c \ra D^{(5) / (4,3)} \big) \nonumber \\
 - & \;   \teta_{1 \dc} \eta_{2 c} \frac{1}{s_{12}} \big( s_{12} [ 1^{\dc} | \hap_2 \hap_5 |Ê2^c \ra B^{(5) / (4,3)} +  [ 1^{\dc} | \hap_2 \hap_5 \hap_3 \hap_1 |Ê2^c \ra C^{(5) / (4,3)} \big)   \nonumber \\
+ & \;  \teta_{2 \dc} \eta_{1 c} \frac{1}{s_{12}} \big( s_{12} [ 2^{\dc} |  \hap_5  \hap_2 |Ê1^c \ra B^{(5) / (4,3)} + [ 2^{\dc} | \hap_1 \hap_3 \hap_5  \hap_2 |Ê1^c \ra C^{(5) / (4,3)} \big) \nonumber \\
+ & \;  \teta_{1 \dc} \eta_{5 c} \big( [ 1^{\dc} | \hap_2 \hap_1 |Ê5^c \ra A^{(5) / (4,3)} + [ 1^{\dc} | \hap_2 \hap_3 |Ê5^c \ra C^{(5) / (4,3)} \big)  \nonumber \\
- &  \; \teta_{5 \dc} \eta_{1 c}  \big( [ 5^{\dc} | \hap_1 \hap_2  |Ê1^c \ra A^{(5) / (4,3)}  + [ 5^{\dc} | \hap_3 \hap_2  |Ê1^c \ra C^{(5) / (4,3)} \big)  \nonumber \\
+ & \; \teta_{5 \dc} \eta_{2 c} \big(  [ 5^{\dc} | \hap_2  \hap_1 |Ê2^c \ra B^{(5) / (4,3)} + [ 5^{\dc} | \hap_3  \hap_1 |Ê2^c \ra C^{(5) / (4,3)} \big) \nonumber \\
- & \; \teta_{2 \dc} \eta_{5 c} \big(  [ 2^{\dc} | \hap_1 \hap_2 |Ê5^c \ra B^{(5) / (4,3)} + [ 2^{\dc} | \hap_1 \hap_3 |Ê5^c \ra C^{(5) / (4,3)} \big)  \, . 
\end{alignat}
Here, the variables $A^{(5) / (4,3)}, B^{(5) / (4,3)}, C^{(5) / (4,3)}$ and $D^{(5) / (4,3)}$ are the coefficients from the PV reduction of the scalar pentagon $I_5$ or box function $I_{4,3}$ respectively. 
For the scalar pentagon, we have 
\begin{alignat}{1}
A^{(5)}  = & \; \Delta^{-1} ( s_{15} s_{13} s_{23} s_{25} + s_{15} s_{25} s_{23}^2 - s_{13} s_{23} s_{25}^2 - s_{13}^2 s_{25}^2 + 2 s_{12} s_{13} s_{25} s_{35}  \nonumber \\ 
& + s_{12} s_{23} s_{35} s_{15} + s_{12} s_{23} s_{25} s_{35} - s_{12}^2 s_{35}^2 )  \nonumber \\
B^{(5)} = & \; \Delta^{-1} s_{15} ( s_{12} s_{23} s_{35} + s_{13} s_{23} s_{25} - s_{12} s_{13} s_{35} - s_{13} s_{23} s_{15} + s_{13}^2 s_{25} - s_{15} s_{23}^2 )  \nonumber \\
C^{(5)} = & \; \Delta^{-1} s_{12} s_{15} (s_{12} s_{35} - s_{15} s_{23} - s_{13} s_{25} - 2 s_{23} s_{25} ) \nonumber \\
D^{(5)} = & \; \Delta^{-1} s_{12} s_{23} ( s_{15} s_{23} + s_{13} s_{25}  - s_{12} s_{35} + 2 s_{15} s_{13} )  
\end{alignat}
whereas for the coefficients of the box integral $I_{4,3}$ we find
\begin{alignat}{1}
A^{(4,3)}  = & \; \Delta^{-1} s_{25} ( s_{13} s_{25} - s_{15} s_{23} - s_{12} s_{35} ) \nonumber \\
B^{(4,3)} = & \; \Delta^{-1} s_{15} ( s_{15} s_{23} - s_{13}  s_{25} - s_{12} s_{35} )  \nonumber \\
C^{(4,3)} = & \; \Delta^{-1} 2 s_{12} s_{15} s_{25}  \nonumber \\
D^{(4,3)} = & \; \Delta^{-1} s_{12} ( s_{12} s_{35} - s_{15} s_{23} - s_{13} s_{25} ) \ . 
\end{alignat}
Here, $\Delta$ is the Gram determinant, and is explicitly given by
\be
\Delta = s_{15}^2 s_{23}^2 + (s_{13} s_{25} - s_{12} s_{35} )^2 - 2 s_{15}
s_{23} ( s_{13} s_{25} + s_{12} s_{35}  ) \, .
\ee
Notice that for the final expression for the amplitude we have to collect the five box integrals $I_{4,i}$ with their respective coefficients which can be obtained by cyclic permutation of the states $(1, \dots, 5)$. Furthermore, we have to include one copy of the pentagon integral with its coefficient. The pentagon coefficient does not possess manifest cyclic symmetry, and each of the five quadruple cuts produces a different looking expression. However, our tests provided below confirm that
the pentagon coefficients have the expected cyclic symmetry.

\subsection{Gluon component amplitude}

In this section we extract from the one-loop five-point superamplitude its component where all external particles are six-dimensional gluons.  
This is useful since, dimensionally reducing this component amplitude  to four dimensions, one can access the gluon MHV and anti-MHV amplitudes of $\cN=4$ SYM.

In order to extract this component  we have to integrate one power of $\eta_i$ and $\teta_i$ for each external state, here denoted by  
$1_{a \da}$, $2_{b \db}$, $3_{c \dc}$, $4_{d \dd}$,  and $5_{e \dot{e} }$. 
Doing this,  one arrives at 
\begin{alignat}{1} \label{5pt-1-loop-YG-amplitude-34cut}
& A_{5;1} \big|_{\mbox{\scriptsize (3,4)-cut}}  \propto  \int \frac{d^6 l}{(2 \pi)^6} \frac{1}{s_{34}} \delta^+ (l_1^2) \delta^+ (l_2^2) \delta^+ (l_3^2) \delta^+ (l_5^2) \frac{ 1 }{(p_3 + l_3)^2  } \\
%\begin{alignat}{1} 
%\label{5pt-1-loop-YG-amplitude-34cut}
%& A_{5;1}   \propto  \frac{1}{s_{34}}\int\!\!\frac{d^6 l}{(2 \pi)^6}  \prod_{i=1}^5 {i\over l_i^2}  \\
 \times \bigg\{ & \frac{1}{s_{12}} [ 1_{\da} | \hap_2 \hal_1 \hap_5 \hap_2 | 1_a \ra \la 2_b 3_c 4_d 5_e \ra [ 2_{\db} 3_{\dc} 4_{\dd} 5_{\dot{e}} ] + \frac{1}{s_{12}} [ 2_{\db} | \hap_1 \hal_1 \hap_5 \hap_1 | 2_b \ra \la 1_a 3_c 4_d 5_e \ra [ 1_{\da} 3_{\dc} 4_{\dd} 5_{\dot{e}} ] \nonumber \\
+ &  \frac{1}{s_{15}} [ 5_{\dot{e}} | \hap_1 \hal_1 \hap_2 \hap_1 | 5_e \ra \la 1_a 2_b 3_c 4_d \ra [ 1_{\da} 2_{\db} 3_{\dc} 4_{\dd}  ] \nonumber \\
- & \frac{1}{s_{12}} [ 2_{\db} | \hap_1 \hal_1 \hap_5 \hap_2 | 1_a \ra \la 2_b 3_c 4_d 5_e \ra [ 1_{\da} 3_{\dc} 4_{\dd} 5_{\dot{e}} ] + \frac{1}{s_{12}} [ 1_{\da} | \hap_2 \hap_5 \hal_1 \hap_1 | 2_b \ra \la 1_a 3_c 4_d 5_e \ra [ 2_{\db} 3_{\dc} 4_{\dd} 5_{\dot{e}} ] \nonumber \\
- & [1_{\da} | \hap_2 \hal_1 | 5_e \ra \la 1_a 2_b 3_c 4_d \ra [2_{\db} 3_{\dc} 4_{\dd} 5_{\dot{e}} ] + [5_{\dot{e}} | \hal_1 \hap_2 | 1_a \ra \la 2_b 3_c 4_d 5_e \ra [ 1_{\da} 2_{\db} 3_{\dc} 4_{\dd} ] \nonumber \\
- & [5_{\dot{e}} | \hal_1 \hap_1 | 2_b \ra \la 1_a 3_c 4_d 5_e \ra [ 1_{\da} 2_{\db} 3_{\dc} 4_{\dd}  ] + [2_{\db} | \hap_1 \hal_1 | 5_e \ra \la 1_a 2_b 3_c 4_d \ra [ 1_{\da} 3_{\dc} 4_{\dd} 5_{\dot{e}} ] \bigg\}  \nonumber \ , 
\end{alignat}
where $l_i$, $i=1, \ldots, 5$ are the five propagators in Figure \ref{generic-pentagon}. 
In the next section we perform the reduction to four dimension of  \eqref{5pt-1-loop-YG-amplitude-34cut}, which will give us important checks on our result.

\subsection{4D limit of the one-loop five-point amplitude} \label{4dconcheck} 

%%%%%%%%%%%%%%%%%%%%%%

An important series of nontrivial consistency checks on our six-dimensional  five-point amplitude at one loop 
can be obtained by performing its   reduction to four dimensions, and comparing  it to the expected form of the one-loop (MHV or anti-MHV) amplitude(s)  directly calculated in four-dimensional $\cN=4$ SYM theory. 

In order to perform  the reduction to four dimensions of various six-dimensional quantities, one can employ the results of 
\cite{Boels:2009bv} (see also \cite{siegel}). 
There, it was found that the solutions to the Dirac equation with the external momenta 
living in a four-dimensional subspace, i.e.~$p = (p^0, p^1, p^2, p^3, 0, 0)$, can be written as 
\be
\label{xxx}
\lambda^A_a =
\left(
\begin{array}{cc}
 0 & \lambda_{\alpha}    \\
 \tlambda^{\dot{\alpha}} & 0
\end{array}
\right), \quad \qquad 
 \tlambda_{A \da} = \left(
\begin{array}{cc}
 0 & \lambda^{\alpha}    \\
- \tlambda_{\dot{\alpha}} & 0
\end{array}
\right)
\ , 
\ee
where $\lambda_{\alpha}$ and $\tlambda_{\dot{\alpha}}$ are the usual four-dimensional spinor variables.
Hence,  the Lorentz invariant, little group covariant quantities $\la i_a | j_{\da} ]$, $ [ i_{\da} | j_a \ra$  become
\be \label{4d-limit-6d-lorentz-invariants}
\la i_a | j_{\da} ] = \left(
\begin{array}{cc}
[i j] & 0     \\
0 & - \la i j \ra
\end{array}
\right), \quad \qquad [ i_{\da} | j_a \ra = \left(
\begin{array}{cc}
 - [ i j ] & 0    \\
0 & \la i j \ra
\end{array}
\right) \ .
\ee
Here, we follow the standard convention of writing the four-dimensional spinor contractions as $\lambda_i^{\alpha} \lambda_{j \alpha} = \la i j \ra$ and $\tlambda_{i \dot{\alpha}} \tlambda_j^{\dot{\alpha}} = [i j]$.

The four-dimensional helicity group is a $\mathsf{U}(1)$ subgroup of the six-dimensional little group which preserves the structure of 
\eqref{xxx} and \eqref{4d-limit-6d-lorentz-invariants}.  
In order to determine  the (four-dimensional) helicity of a certain state in \eqref{6d-superfield},  
a practical way to proceed is as follows. 
Each appearance of a dotted or undotted index equal to 1 (2) contributes 
an amount  of $+1/2$  ($-1/2$) to the total four-dimensional helicity.   
As an example, consider the term $A_{a \dot{a}}$ in \eqref{6d-superfield}. 
States with $(a,\da) = (1, 1) $ correspond, upon  reduction,  
to gluons with positive helicity and states with 
$(a,\da) = (2, 2)$ to gluons of  negative helicity.

In the four-dimensional limit, the six-dimensional spinor brackets become%
\footnote{Note that the our labeling of positive and negative helicities in four dimensions is opposite that in \cite{siegel}.}
 \cite{siegel}
\begin{eqnarray} \label{4product}
-\langle i_{+}|j_{+}]=[ij]=[i_{+}|j_{+}\rangle \ ,  &\qquad  & \langle i_{-}|j_{-}]=\langle ij\rangle=-[i_{-}|j_{-}\rangle\ , \\
\langle i_{-}j_{-}k_{+}l_{+}\rangle = - \langle ij\rangle[kl] \ , &\qquad  & [i_{-}j_{-}k_{+}l_{+}] =  - \langle ij\rangle[kl] \ ,  \nonumber \\
\langle i_{-}j_{+}k_{-}l_{+}\rangle = + \langle ik \rangle[ jl] \ , & \qquad & [i_{-}j_{+}k_{-}l_{+}] =  +\langle ik \rangle [ jl] \ . \nonumber
\end{eqnarray}
In the following we will use these identifications to check the  four-dimensional limits of   (\ref{5pt-1-loop-YG-amplitude-34cut})  for all  MHV helicity assignments of  the external gluons. 
As expected, we will always obtain the expected $\cN=4$ SYM result, i.e.~the appropriate Parke-Taylor MHV prefactor multiplied by a  four-dimensional  one-loop box function.

To begin with, we recall that upon four-dimensional reduction, a six-dimensional scalar pentagon reduces to five different box functions (plus terms vanishing in four dimensions) 
\cite{Bern:1993kr,Bern:1992em,5DPentagon}, 
and hence contributes to the coefficients of the relevant box functions. Schematically,  
\be
\mathcal{C}^{(5)} I_5  +  \mathcal{C}^{(4,3)} I_{4,3} \stackrel{4D}{\longrightarrow} \left[ \mathcal{C}^{(5)} \frac{P^{(4,3)} }{2 s_{12} s_{23} s_{34} s_{45} s_{51}}  +  \mathcal{C}^{(4,3)} \right] I_{4,3}
\ee
where
\be
P^{(4,3)} = s_{12} s_{51} ( s_{12} s_{23} - s_{12} s_{51} - s_{23} s_{34} - s_{34} s_{45} + s_{45} s_{51} )
\ , 
\ee
when going to four dimensions. Hence, upon dimensional reduction the coefficients of the PV reduction become 
\begin{alignat}{1}
A \rightarrow &  - \frac{ s_{12} s_{15} - s_{15} s_{45} + s_{34} s_{45}}{2 s_{23} s_{34} s_{45}}, \\
 B \rightarrow & - \frac{ s_{15} s_{12} - s_{15} s_{45} }{2 s_{23} s_{34} s_{45} }, \nonumber \\
C \rightarrow & - \frac{ s_{12} s_{15}}{2 s_{23} s_{34} s_{45} }, \nonumber \\
D \rightarrow & - \frac{s_{12} }{2 s_{34} s_{45} } \nonumber \ .
\end{alignat}
Let us now discuss specific helicity assignments. We  start by considering the configuration  
$(1^-, 2^-, 3^+, 4^+, 5^+)$. In this case,   after  PV reduction only the third term in (\ref{5pt-1-loop-YG-amplitude-34cut}) 
is non-vanishing. Hence, we have to consider the four-dimensional limit of
\be
\frac{1}{s_{3 4} s_{15}} \left( [ 5_{\dot{e}} | \hap_1 \hap_3 \hap_2 \hap_1 | 5_e \ra C + [ 5_{\dot{e}} | \hap_1 \hap_5 \hap_2 \hap_1 | 5_e \ra D \right)   \la 1_a 2_b 3_c 4_d \ra [ 1_{\da} 2_{\db} 3_{\dc} 4_{\dd}  ] \,.\label{eq:nonzero12}
\ee
Upon dimensional reduction, the resulting contribution  is
\be
\label{5p1lreduction}
{s_{15} s_{12}\over 2} \, \frac{ \la 12 \ra^3}{\la 23 \ra \la 34 \ra \la 45 \ra \la 51 \ra} \,.
\ee
Given the relation between the scalar box functions  $F_4$ and the corresponding box integrals,
$I_4= 2\, F/ (s_{12} s_{15}) $, it is immediate  to see that the kinematic factors  in (\ref{5p1lreduction}) cancel and the final result is the anticipated one:
\be
\frac{\la 1 2 \ra^4}{\la 1 2 \ra \la 2 3 \ra \la 3 4 \ra \la 4 5 \ra \la 5 1 \ra}\,.
\ee
The fact that the form of the one-loop five-point amplitude upon reduction to four dimensions is precisely the well-known result is an expected, though  highly non-trivial, outcome.

As mentioned above,  we have performed checks  for all external helicity configurations, finding in all cases agreement with the expected four-dimensional result. We would like to highlight a particularly stringent test, namely that  
corresponding  to the helicity configuration 
 $( 1^+, 2^+, 3^-, 4^-, 5^+)$, where all terms in  \eqref{5pt-1-loop-YG-amplitude-34cut} contribute to the four-dimensional reduction.

A final comment is in order here. 
It is known that collinear and soft limits put important constraints on tree-level and loop amplitudes in any gauge theory and in gravity. In six dimensions, the lack of infrared divergences makes loop level factorisation trivial, similarly to what happens to four-dimensional gravity because of its improved infrared behaviour compared to four-dimensional Yang-Mills theory amplitudes. Therefore, the  factorisation properties we derive below from tree-level amplitudes 
will apply unmodified to one-loop amplitudes.

We now consider again the five-point amplitude \eqref{5pttree} derived in 
 \cite{cheung}, and  take the soft limit where 
$p_1 \to 0$.   A short calculation shows that 
\beq
\label{5pttreesoft} 
A_{5; a \dot{a} \ldots }^{(0)}  \to S_{a \dot{a}} (5, 1, 2) A_{4; \ldots }^{(0)}
\ ,
\eeq
where we find, for the six-dimensional soft function,
\beq
\label{soft6da}
S_{a \dot{a}} (i, s, j) \ = \ { \langle s_a |\hat{p}_j \hat{p}_i |  s_{\dot{a} }]  \over s_{is} s_{sj} 
}
\ .
\eeq
In \eqref{5pttreesoft}  the dots stand for the little group indices of the remaining particles in the amplitude. 
Using the results in this section, it is also immediate to check  that \eqref{soft6da} reduces, in the four-dimensional limit, to the expected soft functions of \cite{mp}. 
As a final test on our five-point amplitude we have checked that the soft limits where legs 1, 2 or 5 become soft are all correct.

This provides an exhaustive set of checks of our result for the six-dimensional five-point superamplitude.

%%%%%%%%%%%%%%%%%%%%%%%%%%%%%%%%%

\newpage

\section*{Acknowledgements}

It is a pleasure to thank   Niklas Beisert, David Berman, Bill Spence, Dan Thompson and Gang Yang for interesting discussions.
This work was supported by the STFC under a Rolling Grant  ST/G000565/1.
GT is supported by an EPSRC Advanced Research Fellowship EP/C544242/1.

\appendix

\section{Notation and conventions}
\label{appA}

In this appendix we collect some details on our normalisations and conventions. 
\newline
The total antisymmetric $\mathsf{SU}(2)$-invariant tensors are given by
\begin{equation}
\epsilon_{a b} = 
\left(
\begin{array}{cc}
0 & -1     \\
1 & 0   
\end{array}
\right),
\hspace{1cm}
\epsilon^{a b} = 
\left(
\begin{array}{cc}
0 & 1     \\
-1 & 0   
\end{array}
\right).
\end{equation}
The Grassmann integration  measure  is defined as  
$ d^2 \eta =  (1/2) d \eta^a d \eta_{ a}  =  d \eta_{ 2} d \eta_{1}$,   
such that
\begin{alignat}{1}
\int\!\!d^2 \eta \; \big[ \lambda^{A a} \eta_a \;  \lambda^{B b} \eta_b \big]  & = - \left( \lambda^{A a} \lambda^B_a \right) \, . 
\end{alignat}
The Clebsch-Gordan symbols are normalised as 
\be
\tilde\sigma_{\mu}^{AB}:= {1\over2} \eps^{ABCD}\, \sigma_{\mu, CD}
\, ,
\eeq
with 
\beq
\Tr ( \sigma^\mu \tilde\sigma^\nu) \, = \, 4\, \eta^{\mu \nu}
\ . 
\ee
Using these relations, the scalar product of two vectors $p$ and $q$ can equivalently be expressed as 
\beq 
\label{dotpr}
p\cdot q = - {1\over 4} p^{AB}q_{AB} \ = \ -{1\over 8} \eps_{ABCD} p^{AB} q^{CD}
\ , 
\eeq
where $p^{AB} := p^\mu \tilde{\sigma}_{\mu}^{AB} $ and $p_{AB} := p^\mu \sigma_{\mu, AB} $. \newline
\begin{figure}[t]
\begin{center}
\includegraphics[width=7cm]{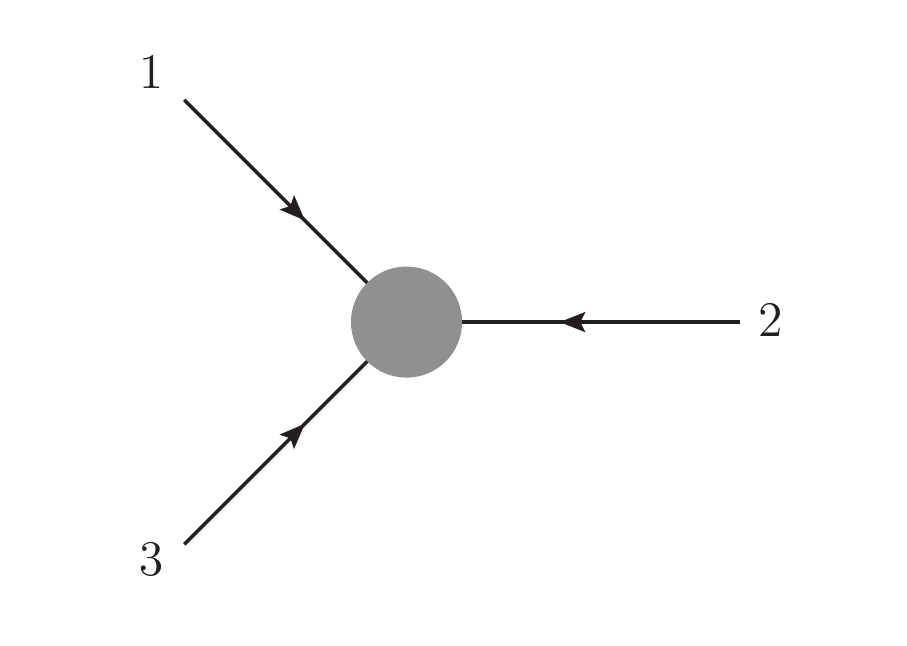}
\caption{\it A generic three-point vertex with all momenta defined as incoming.}
\label{generic-3pt-vertex}
\end{center}
\end{figure}
Momentum conservation for three-point amplitudes implies  that $p_{i}\cdot p_{j}=0$, $i, j=1,2,3$. In six dimensions, this condition is equivalent to \cite{cheung} 
\beq
\label{zerocond}
\mathrm{det}\langle i|j]_{a\dot{a}}\, = \, 0
\, 
\eeq
where $\lambda_{ia}^{A}\tilde{\lambda}_{Aja}:=\langle i_{a}|j_{\dot{a}}]$
and we used $p_{i}^{AB}=\lambda_{ia}^{A}\lambda_{i}^{Ba}$ and $p_{iAB}=\tilde{\lambda}_{i A}^{\da}  \tilde{\lambda}_{i B \da }$.
Hence, \eqref{zerocond} allows to recast the matrix $\langle i_{a}|j_{\dot{a}}]$ as  a product of two spinors, as 
\cite{cheung}
\beq
\label{tbtm}
\langle i_{a}|j_{\dot{b}}]= (-)^{\cP_{ij}}u_{ia}\tilde{u}_{j\dot{b}}
\ , 
\eeq
where we choose $(-)^{\cP_{ij}} = +1$ for $(i,j) =(1,2), (2,3), (3,1)$, and $-1$ for   $(i,j) =(2,1), (3,2), (1,3)$. Hence, for a generic three-point vertex with all momenta defined to be incoming (see Figure \ref{generic-3pt-vertex}) we have a positive sign when rewriting Lorentz contracted spinor combinations in a clockwise ordering. 
\newline

\section{Supersymmetry invariance of the three-point superamplitudes} 
\label{appD}

Here we provide an explicit proof of the fact that the three-point superamplitude \eqref{tree-3pt-super}
 is supersymmetric. We choose to decompose each variable $\eta_i$  as
\beq
\eta_{i}^{a}=u_{i}^{a}\eta_{i}^{\parallel}+w_{i}^a\eta_{i}^{\perp}
\ , 
\eeq
which is a convenient choice since $u_i^a w_{i a} =1$.  We also notice that, using this decomposition, we can recast  the quantities $W$ and $\tilde{W}$ defined in \eqref{w} entering the expression of the three-point superamplitude, as 
\beq
\label{WW2}
W \ = \sum_{i=1}^3 \eta_{i}^{\parallel} \ , \qquad 
\tilde{W} \ = \sum_{i=1}^3 \tilde{\eta}_{i}^{\parallel}
\ .
\eeq
The supersymmetry generators can then be written as
\beq
\label{ab3}
Q^{A}=\sum_{i}\lambda_{i}^{A a} u_{i a} \eta_{i}^{\parallel}+\sum_{i}\lambda_{i}^{A a}w_{i a} \eta_{i}^{\perp}\, . 
\eeq
A direct consequence of six-dimensional momentum conservation is the fact that the quantities $\lambda_{i}^{A a}u_{i a}$ are $i$-independent, 
therefore we can rewrite  \eqref{ab3} in several equivalent ways, one of which is
\beq
\label{qrew}
Q^{A}=(\lambda_{1}^{A a} u_{1 a} )W + ( \lambda_{1}^{A a} w_{1 a} )(\eta_{2}^\perp-\eta_{1}^\perp)+(\lambda_{2}^{A a} w_{2 a} ) (\eta_{3}^\perp-\eta_{1}^\perp)
\ ,
\eeq
where $W$ is given in \eqref{WW2}, and the constraint on the $w$'s \eqref{wconstraint} is used.
Using the  decomposition \eqref{qrew} it is very easy to prove that $Q^A A_3 = 0$.  
To this end, we first observe that the presence of a factor $\delta (W) \delta (\tilde{W})$ in \eqref{tree-3pt-super} effectively removes the first term  from the expression of \eqref{qrew}, and we are left to prove that  
$Q^A_\perp := (\lambda_{1}^{A a} w_{1 a} )( \eta_{2}^\perp-\eta_{1}^\perp)+(\lambda_{2}^{A a} w_{2 a} )(\eta_{3}^\perp-\eta_{1}^\perp) $ annihilates the amplitude. 
Specifically, we will show that  
\beq
\label{eqab}
Q^A_\perp \, \Big[ \delta(Q^A) \delta(\tilde{Q}_A ) \Big]^2\ = \ 0
\, . 
\eeq
To begin with, we observe that 
\beqa
\label{eq:delta argument}
 \delta(Q^A) \delta(\tilde{Q}_A )  & = &  
 \sum_{i,j=1}^3 \la i_a | j_{\dot{a}} ]  \eta_i^a {\tilde{\eta}}_j^{\da}
= 
 \sum_{i,j=1}^3 (-)^{\cP_{ij}} u_{ ia}  \tilde{u}_{j \dot{a}}  (1 - \delta_{ij}) \eta_i^a {\tilde{\eta}}_j^{\da}
 \nonumber \\
&=& \sum_{i,j=1}^3  (-)^{\cP_{ij}} (1 - \delta_{ij}) \eta_i^\perp {\tilde{\eta}}_j^{\perp}
\nonumber \\
&=&
\eta^{\perp}_1 \tilde{\eta}^{\perp}_2 - \eta^{\perp}_1 \tilde{\eta}^{\perp}_3 - \eta^{\perp}_2 \tilde{\eta}^{\perp}_1 + \eta^{\perp}_2 \tilde{\eta}^{\perp}_3 + \eta^{\perp}_3 \tilde{\eta}^{\perp}_1 - \eta^{\perp}_3 \tilde{\eta}^{\perp}_2
\ , 
\eeqa 
where we have used \eqref{tbtm}. Using \eqref{eq:delta argument}, one then finds (we drop the superscript $\perp$ in the following)
\begin{alignat}{1}
\nonumber
%\hspace{-0.7cm}
\Big[ \delta(Q^A) \delta(\tilde{Q}_A ) \Big]^2 \! \! = & -\eta_{1} \tilde{\eta}_{2} \eta_{2} \tilde{\eta}_{1} + \eta_{1} \tilde{\eta}_{2} \eta_{2} \tilde{\eta}_{3} + \eta_{1} \tilde{\eta}_{2} \eta_{3} \tilde{\eta}_{1} + \eta_{1} \tilde{\eta}_{3} \eta_{2} \tilde{\eta}_{1} - \eta_{1} \tilde{\eta}_{3} \eta_{3} \tilde{\eta}_{1} + \eta_{1} \tilde{\eta}_{3} \eta_{3} \tilde{\eta}_{2}
\nonumber \\
 &   - \eta_{2} \tilde{\eta}_{1} \eta_{1} \tilde{\eta}_{2} + \eta_{2} \tilde{\eta}_{1} \eta_{1} \tilde{\eta}_{3} + \eta_{2} \tilde{\eta}_{1} \eta_{3} \tilde{\eta}_{2} + \eta_{2} \tilde{\eta}_{3} \eta_{1} \tilde{\eta}_{2} + \eta_{2} \tilde{\eta}_{3} \eta_{3} \tilde{\eta}_{1} - \eta_{2} \tilde{\eta}_{3} \eta_{3} \tilde{\eta}_{2}
 \nonumber \\
 &   + \eta_{3} \tilde{\eta}_{1} \eta_{1} \tilde{\eta}_{2} - \eta_{3} \tilde{\eta}_{1} \eta_{1} \tilde{\eta}_{3} + \eta_{3} \tilde{\eta}_{1} \eta_{2} \tilde{\eta}_{3} + \eta_{3} \tilde{\eta}_{2} \eta_{1} \tilde{\eta}_{3} + \eta_{3} \tilde{\eta}_{2} \eta_{2} \tilde{\eta}_{1} -\eta_{3} \tilde{\eta}_{2} \eta_{2} \tilde{\eta}_{3}
  \ . 
\end{alignat}
Next, we calculate
\beqa
\eta_1 \Big[ \delta(Q^A) \delta(\tilde{Q}_A ) \Big]^2 &=&
2\eta^{1}\eta^{2}\eta^{3}(\tilde{\eta}^{1}\tilde{\eta}^{3}-\tilde{\eta}^{1}\tilde{\eta}^{2}-\tilde{\eta}^{2}\tilde{\eta}^{3})
\ , 
\eeqa
and furthermore we find that 
\beq
\label{eqcon}
\eta_2 \Big[ \delta(Q^A) \delta(\tilde{Q}_A ) \Big]^2 \, = \,  
\eta_3 \Big[ \delta(Q^A) \delta(\tilde{Q}_A ) \Big]^2 \, = \, 
\eta_1 \Big[ \delta(Q^A) \delta(\tilde{Q}_A ) \Big]^2\, . 
\eeq
Inspecting the form of  $Q^{A}$  in \eqref{qrew} and using \eqref{eqcon},
we conclude that \eqref{eqab} holds, and therefore the three-point superamplitude is invariant under supersymmetry.

\section{Useful spinor identities in six dimensions}
\label{appB}

In this appendix we collect identities between six-dimensional spinor variables that we have frequently used in the  calculations presented in this paper.

We begin by quickly stating two basic relations for three-point spinors $u_{i, a}$ and $w_{i, a}$. For a general three-point amplitude in six dimensions we have \cite{cheung}
\begin{alignat}{1}
& u_i^a | i_a \ra = u_j^b | j_b \ra \ , \qquad \qquad 
 \tu_i^{\da} | i_{\da} ] = \tu_j^{\db} | j_{\db} ] \ .
% \nonumber \\
%& w_1^a | 1_a \rangle + w_2^b | 2_b \rangle + w_3^c | 3_c \rangle = 0 \ , \nonumber \\
%& \tw_1^{\da} | 1_{\da} ] + \tw_2^{\db} | 2_{\db} ] + \tw_3^{\dc} | 3_{\dc} ]  = 0
\end{alignat}
We also have  the constraints   \eqref{wconstraint} on the $w$'s  and their $\tilde{w}$  counterparts, which are essentially a consequence  of momentum conservation.

\begin{figure}[t]
\begin{center}
\includegraphics[width=8cm]{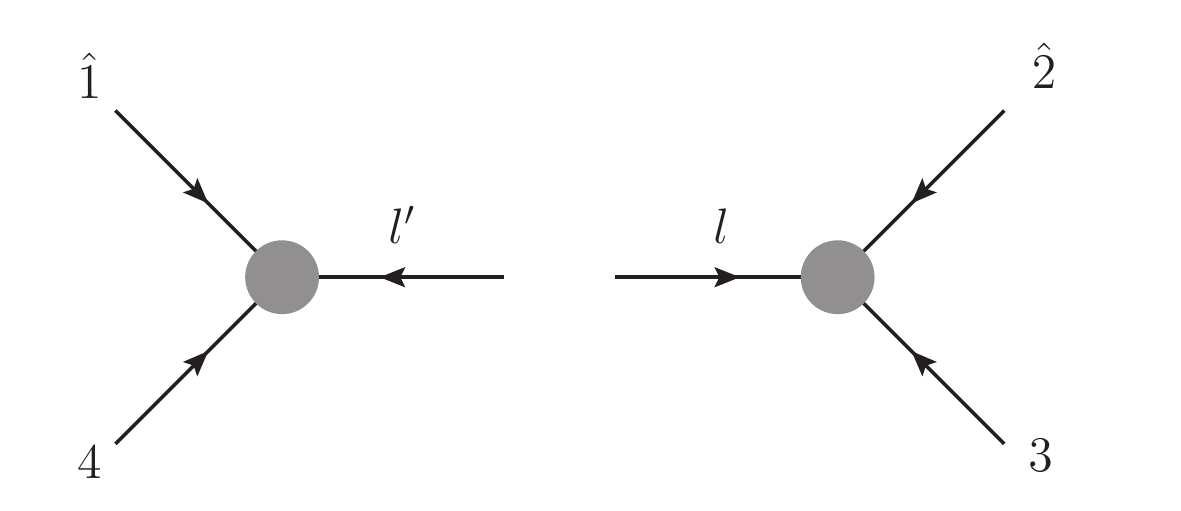}
\caption{\it The recursive construction of a four-point tree-level amplitude. The shifted legs are $1$ and $2$ and we have $l' = -l$ for the internal propagator.}
\label{4pt-recursive-diagram}
\end{center}
\end{figure}

Next, we make use of relations between two three-point amplitudes, connected by an internal propagator, just as in the BCFW construction of the four-point amplitude. We give a pictorial representation of this  in Figure \ref{4pt-recursive-diagram}. 
We have defined the internal momenta $l$ and $l'$ to be incoming for the three-point amplitudes, giving the relation $l' = - l$. Since six-dimensional momenta are products of two spinors we can define
\be
| l_i' \rangle = i | l_i \rangle \ ,  \qquad \qquad   | l_i \rangle = (-i) | l_i' \rangle\ , 
\ee
and similarly for $\tlambda$-spinors. Also note that we can normalise the spinors $u_a, w_b$ of one three-point subamplitudes in Figure \ref{4pt-recursive-diagram} such that they are related to the spinors of the other subamplitude, yielding (see Appendix \ref{appB2})
\be \label{normalisation-w-u}
w_{l_i' a} = \frac{u_{l_i a }}{\sqrt{-s}} \ , \qquad \qquad  w_{l_i a} = - \frac{u_{l_i' a} }{\sqrt{-s}} \ . 
\ee
Similar expressions hold for the spinors $\tu_{\da}, \tw_{\db}$. In the following we will be discussing several relations in the cases of the four- and five-point amplitudes. 

\subsection{Product of two $u$-spinors}
\label{appB1}

In the calculation of the five-point cut-expression we encounter $u$-spinors belonging to the same external state and would like to remove them from the expression. Consider the object $u_{i a} \tu_{i \da}$ with states $p_i$ and $p_j$  belonging to the same three-point amplitude. We can write \cite{siegel}
\begin{alignat}{1} \label{u-tu-expression}
u_{i a} \tu_{i \da} & = u_{i a}  \tu_{i \db}  \delta^{\db}_{\da} =  u_{i a}  \tu_{i \db} \frac{ \langle P_b | i^{\db} ] }{ \langle P_b | i^{\da}] }  \nonumber \\
& =  u_{i a}  \tu_{i \db}  \langle P_b | i^{\db} ]  \langle P_b | i^{\da} ]^{-1} =   u_{i a}  \tu_{i \db}   [ i^{\db} |  P_b \ra \la P^b | i_{\da} ] \frac{ 1 }{ s_{i P} } \nonumber \\
& = u_{i a} \tu_{j \db} [ j^{\db} |  P_b \ra \la P^b | i_{\da} ] \frac{ 1 }{ s_{i P} } = (-)^{\cP_{ij}} \la i_a | j_{\db} ] [ j^{\db} |  P_b \ra \la P^b | i_{\da} ] \frac{ 1 }{ s_{i P} }  \nonumber \\
& =  \frac{ (-)^{\cP_{ij}} }{ s_{i P} }  \la i_a | \hap_{j} \hap_{P} | i_{\da} ]
\ , 
\end{alignat}
where we have $(-)^{\cP_{ij}} = +1$ for clockwise ordering of the states $(i,j)$ for the three-point amplitude. Also, $P$ is an arbitrary momentum. By the same series  of manipulations we can show that 
\begin{alignat}{1} \label{u-tu-expression-2}
u_{i a} \tu_{i \da} & = u_{i b}  \tu_{i \da}  \delta^{b}_{a} = \frac{ (-)^{\cP_{ji}} }{ s_{i P} }  \la i_a | \hap_{P} \hap_{j} | i_{\da} ] \ . 
\end{alignat}
Note that the difference between (\ref{u-tu-expression}) and (\ref{u-tu-expression-2}) is just a sign since $(-)^{\cP_{ji}} = - (-)^{\cP_{ji}}$.

\subsection{The relation $ w_l \cdot  w_{l'} \; \tw_l \cdot \tw_{l'} = - s^{-1}_{i j}$}
\label{appB2}

Here we provide an expression for the contraction between $w$- and $\tw$-spinors of two three-point amplitudes, connected by an internal propagator, originally encountered in the recursive calculation of the four-point tree amplitude in \cite{cheung}. 

We start with expression \ref{u-tu-expression} and choose $i=1, j=4$ and $P=2$, following Figure \ref{4pt-recursive-diagram}. This yields
\be \label{u-u-tu-tu-expr1}
u_{1 a} \tu_{1 \da} s_{\hone \htwo} =  - \la \hone_a | \hap_{4} \hap_{\htwo} | \hone_{\da} ] \ .
\ee
However, we can also write
\begin{alignat}{1} \label{u-u-tu-tu-expr2}
\la \hone_a | \hap_{4} \hap_{\htwo} | \hone_{\da} ] & = - u_{\hone a} \tu_{4}^{\dd} [ 4_{\dd} |Ê\htwo^b \ra \la \htwo_b | \hone_{\da} ] = - u_{\hone a} \tu_{l'}^{\dd} [ l'_{\dd} |Ê\htwo^b \ra \la \htwo_b | \hone_{\da} ] \nonumber \\
& = (-i)  u_{\hone a} \tu_{l'}^{\dd} [ l_{\dd} |Ê\htwo^b \ra \la \htwo_b | \hone_{\da} ] = i u_{\hone a} \tu_{l'}^{\dd} \tu_{l \dd} u_{\htwo}^b \la \htwo_b | \hone_{\da} ] \nonumber \\
& =  i u_{\hone a} \tu_{l'}^{\dd} \tu_{l \dd} u_{l}^b \la l_b | \hone_{\da} ] = u_{\hone a} \tu_{l'}^{\dd} \tu_{l \dd}  u_{l}^b \la l'_b | \hone_{\da} ] \nonumber \\
& = - u_{\hone a} \tu_{l'}^{\dd} \tu_{l \dd}  u_{l}^b u_{l' b} \tu_{\hone \da} = - u_{\hone a} \tu_{\hone \da} \tu_{l'} \cdot \tu_{l} \; u_{l} \cdot u_{l' } \ .
\end{alignat}
Comparing (\ref{u-u-tu-tu-expr1}) and (\ref{u-u-tu-tu-expr2}) we conclude
\be \label{u-u-tu-tu-final}
\tu_{l'} \cdot \tu_{l} \; u_{l'} \cdot u_{l } = - s_{1 2} 
\ , 
\ee
since $s_{\hone \htwo}  = s_{12}$. Now we express the contractions of $u$-spinors in terms of $w$-spinors. As discussed in \cite{cheung} we can deduce from (\ref{u-u-tu-tu-final}) that 
\be \label{u-w-zero-combinations}
u_l \cdot w_{l'} = \tu_{l} \cdot \tw_{l'} = w_l \cdot u_{l'} = \tw_{l} \cdot \tu_{l'} = 0 \ , 
\ee
by using the redundancy of the $w$-spinors under a shift $w_{l a} \rightarrow w_{l a} + b_l u_{l a}$. Exploiting the defining relation between a spinor $u_l$ and its inverse $w_l$ and multiplying by $u_{l', a}$ and $w_{l', b}$ we have
\be
u_l^a u_{l', a} w_l^b w_{l', b} - u_l^b w_{l', b} w_l^a w_{l',a} = u_{l', a} w_{l', b} \epsilon^{a b} \ .
\ee
Now, the second term on the RHS vanishes as stated in (\ref{u-w-zero-combinations}).  Since $u_l \cdot u_{l'} \neq 0$, we have the relation
\be
u_{l} \cdot u_{l'}  \; w_{l} \cdot w_{l'}  = 1 \; \; \Leftrightarrow \; \;  u_{l} \cdot u_{l'} = \frac{1}{w_{l} \cdot w_{l'} } \ .
\ee
From this we can deduce that a spinor $w_{l a} / w_{l' b}$ is related to the spinor $u_{l' a} / u_{l b}$, respectively, and we can choose to normalise as in (\ref{normalisation-w-u})
\be 
w_{l_i' a} = \frac{u_{l_i a }}{\sqrt{-s_{i j}}} \ , \qquad \qquad  w_{l_i a} = - \frac{u_{l_i' a} }{\sqrt{-s_{i j}}}\ . 
\ee

\subsection{Spinor identities for the one-loop five-point calculation}
\label{appB3}

Here we would like to outline some steps of the calculation which takes us from (\ref{5pt-quad-result-grassmann-int}) to (\ref{5pt-quad-result-prePV}). 

The basic idea is to express the result of the Grassmann integration as a sum of coefficients of factors $\teta_{i \dc} \eta_{j c}$ with $i, j = 1, 2, 5$ for the $(3,4)$-cut. It is then a matter of algebra to rewrite the coefficient of $\teta_{i \dc} \eta_{j c}$ in such a way that any dependence on the three-point quantities $w_{l_i}$, $w_{l_i'}$ and their counterparts in $\teta_{l_i}$ is removed. In the following we provide some explicit terms as examples.

Let us consider one of the terms of the product in (\ref{5pt-quad-result-prePV}), e.g.
\begin{alignat}{1} \label{term_eta1_teta1}
\teta_{1 \dc} \eta_{1 c} & \bigg\{ \la 1^c | l_3 ] \cdot \tw_{l_3'} \tw_{l_2'} \cdot [ l_2 |Ê\hal_1 | l_4 ] \cdot \tw_{l_4} -  \la 1^c | l_2 ] \cdot \tw_{l_2'} \tw_{l_3'} \cdot [ l_3 |Ê\hal_1 | l_4 ] \cdot \tw_{l_4} \bigg\} \nonumber \\
\times & \bigg\{  [ 1^{\dc} |Êl_3 \ra \cdot w_{l_3'} w_{l_2'} \cdot \la l_2 |Ê\hal_1 | l_4 \ra \cdot w_{l_4} - [ 1^{\dc} |Êl_2 \ra \cdot w_{l_2'} w_{l_3'} \cdot \la l_3 |Ê\hal_1 | l_4 \ra \cdot w_{l_4} \bigg\} \ .
\end{alignat}
Firstly,  we realise is that the two factors in the brackets antisymmetrise among themselves. This can be seen by applying the normalisation relations for the $w$-spinors of the internal lines
\begin{alignat}{1}
 [ 1^{\dc} |Êl_3 \ra \cdot w_{l_3'} w_{l_2'} \cdot \la l_2 |Ê\hal_1 | l_4 \ra \cdot w_{l_4} & =  [ 1^{\dc} |Êl_3 \ra \cdot w_{l_3'} \frac{u_{l_2}^a}{\sqrt{-s_{12}}} \la l_{2 a} |Ê\hal_1 | l_4 \ra \cdot w_{l_4} \nonumber \\
 & =  \frac{1}{\sqrt{-s_{12}}} u_{l_3'}^a w_{l_3'}^b [ 1^{\dc} |Êl_{3 b} \ra  \la l_{3 a}' |Ê\hal_1 | l_4 \ra \cdot w_{l_4}
\end{alignat}
where a similar relation is used for the second term of each bracket factor. Since 
\be
u_{l_3'}^a w_{l_3'}^b - u_{l_3'}^b w_{l_3'}^a = \epsilon^{a b}\ , 
\ee
we can write (\ref{term_eta1_teta1}) as
\begin{alignat}{1}
& \teta_{1 \dc} \eta_{1 c} \big( \frac{1}{\sqrt{-s_{12}}} \big)^2 [ 1^{\dc} |Êl_{3 b} \ra \epsilon^{a b} \la l_{3 a}' |Ê\hal_1 | l_4 \ra \cdot w_{l_4} \;  \la 1^c | l_{3 \db} ] \epsilon^{\da \db} [ l_{3 \da}' |Ê\hal_1 | l_4 ] \cdot \tw_{l_4} \nonumber \\
= \; &  \teta_{1 \dc} \eta_{1 c} \frac{i^2}{- s_{12}}  [ 1^{\dc} |Êl_{3}^a \ra \la l_{3 a} |Ê\hal_1 | l_4 \ra \cdot w_{l_4} \; \la 1^c | l_{3}^{\da} ]  [ l_{3 \da} |Ê\hal_1 | l_4 ] \cdot \tw_{l_4} \nonumber \\
= \; & \teta_{1 \dc} \eta_{1 c} \frac{(-1)}{- s_{12}} [ 1^{\dc} | \hal_3 \hal_1 | l_4 \ra \cdot w_{l_4} \; \la 1^c | \hal_3Ê\hal_1 | l_4 ] \cdot \tw_{l_4} \nonumber \\
= \; & \teta_{1 \dc} \eta_{1 c} \frac{1}{ s_{12}} [ 1^{\dc} | \hap_2 \hal_1 | l_4 \ra \cdot w_{l_4} \; \tw_{l_4} \cdot [ l_4 | \hal_1 \hap_2 |Ê1^2 \ra \ , 
\end{alignat}
where we have used momentum conservation at the second corner, $l_3 = l_2 + p_2 = l_1 + p_1 + p_2$, in the last line. 

The next step is to remove the dependence on the $w$-spinors.  The following relation holds:
\begin{alignat}{1}
\hal_1 | l_{4 a} \ra w_{l_4}^a \tw_{l_4}^{\da} [ l_{4 \da} | \hal_1 & = \hap_5 |  l_{4 a} \ra w_{l_4}^a \tw_{l_4}^{\da} [ l_{4 \da} | \hap_5 =  \hap_5 |  l_{1 a}' \ra w_{l_1'}^a (-1)^2  \tw_{l_1'}^{\da} [ l_{1 \da}' | \hap_5 \nonumber \\
& = \big( \frac{1}{\sqrt{-s_{15}}} \big)^2 \hap_5 |  l_{1 a}' \ra u_{l_1}^a   \tu_{l_1}^{\da} [ l_{1 \da}' | \hap_5 = \frac{1}{s_{15}} \hap_5 |  1_{ a} \ra u_{1}^a   \tu_{1}^{\da} [ 1_{\da} | \hap_5 \ .
\end{alignat}
Using the result of (\ref{u-tu-expression}) we arrive at the following string of momenta,
\be
\teta_{1 \dc} \eta_{1 c} \frac{1}{ s_{12} s_{15} s_{1 P} } [ 1^{\dc} | \hap_2 \hap_5 \hap_1 \hal_1 \haP \hap_1 \hap_5 \hap_2 |Ê1^c \ra \ .
\ee
Choosing now $P = p_5$, after  some rearrangement of the momenta we arrive at
\be
\teta_{1 \dc} \eta_{1 c} \frac{1}{ s_{12} s_{15} } [ 1^{\dc} | \hap_2 \hap_5 \hap_1 \hal_1 \hap_5 \hap_2 |Ê1^c \ra \ .
\ee
This expression can be further simplified as follows: Since $l_1 = p_5 + l_4$ we have
\be
\hal_1 \hap_5 = \hal_1 (\hal_1 - \hal_4) = - (\hal_4 + \hap_5) \hal_4 = - \hap_5 \hal_1 \ .
\ee
Permuting now the string of external momenta the final result for the coefficient becomes
\be
- \teta_{1 \dc} \eta_{1 c} \frac{1}{ s_{12} } [ 1^{\dc} | \hap_2 \hap_5 \hal_1 \hap_2 |Ê1^c \ra = \teta_{1 \dc} \eta_{1 c} \frac{1}{ s_{12} } [ 1^{\dc} | \hap_2 \hal_1 \hap_5 \hap_2 |Ê1^c \ra
\ee
by rearranging the order of $\hap_5$ and $\hal_1$ again. 

This algebraic procedure can then be similarly repeated to simplify  all the other coefficients in the cut expression (\ref{5pt-quad-result-grassmann-int}).

\section{PV reduction of the linear pentagon $I_{5, l_1}^{\mu}$}
\label{appC}

We have found in Section \ref{5555} that the one-loop five-point superamplitude can be expressed in terms of just a single function, namely a  linear pentagon,  
\be 
I^{\mu}_{5, l_1} (1, \dots, 5) := \int \frac{d^D l}{(2 \pi)^D}  \frac{l_1^{\mu}}{ l_1^2 l_2^2 l_3^2 (p_3 +l_3 )^2 l_5^2} \, .
\ee
This can be decomposed on  a basis of four independent momenta, as
\be
I^{\mu}_{5, l_1} (1, \dots, 5) = A p_1^{\mu} + B p_2^{\mu} + C p_3^{\mu} + D p_5^{\mu} \, . 
\ee
The choice of the basis vectors is most convenient one due to the kinematical structure of the cut expression in (\ref{5pt-quad-result-prePV}). Contracting with the basis momenta yields
\begin{alignat}{1} \label{PV-I43-equ-syst}
 2 p_1 \cdot I_{5, l_1} = & \int \frac{d^D l}{(2 \pi)^D} \frac{2 p_1 \cdot l_1}{\prod_{i=1}^5 l_i^2}  = I_{4,1} - I_{4,5} \stackrel{!}{=} B s_{12} + C s_{13} + D s_{1 5} \ , 
 \nonumber \\
2 p_2 \cdot I_{5, l_1} = & \int \frac{d^D l}{(2 \pi)^D} \frac{2 p_2 \cdot l_1}{\prod_{i=1}^5 l_i^2}  = I_{4,2} - I_{4,1} - s_{1 2} I_5 \stackrel{!}{=} A s_{12} + C s_{23} + D s_{2 5} \ , 
\nonumber \\
2 p_3 \cdot I_{5, l_1} = & \int \frac{d^D l}{(2 \pi)^D} \frac{2 p_3 \cdot l_1}{\prod_{i=1}^5 l_i^2}  = I_{4,3} - I_{4,2} - (s_{1 2} + s_{23}) I_5 \stackrel{!}{=} A s_{13} + B s_{23} + D s_{3 5} \ , 
\nonumber \\
2 p_5 \cdot I_{5, l_1} = & \int \frac{d^D l}{(2 \pi)^D} \frac{2 p_5 \cdot l_1}{\prod_{i=1}^5 l_i^2}  = I_{4,4} - I_{4,4} \stackrel{!}{=} A s_{15} + B s_{25} + C s_{3 5} \ .
\end{alignat}
Solving the set of linear equations in (\ref{PV-I43-equ-syst}), one obtains  the desired coefficients $A,B, C$ and $D$, used in Section \ref{pvred}. 

%Since the linear pentagon was derived from a certain cut-diagram, we can only derive the coefficients for the scalar %pentagon integral ($A^{(5)}, B^{(5)}, C^{(5)}, D^{(5)})$ and the scalar box integral $(A^{(4,3)}, B^{(4,3)}, C^{(4,3)}, %D^{(4,3)})$ from this set of equations.

%%%%%%%%%%%%%%%%%%%%%%%%%%%%%%%%%%%%%

\section{Reduction to four dimensions of six-dimensional Yang-Mills amplitudes}
\label{appE}

In this appendix we consider the four-dimensional limit of the four- and five-point tree-level amplitudes in pure Yang Mills theory,  and  provide detailed information of how the calculations of  Section \ref{4dconcheck} are carried out.

We begin with the four-point amplitude of \cite{cheung}, given by \eqref{tree-4pt}, and reduce down to a four-dimensional amplitude with 
 helicities $(1^-,2^-,3^+,4^+,5^+)$. Using (\ref{4product}),   the four-dimensional reduction of  (\ref{tree-4pt})
yields
\begin{flalign}
A_{4}^{(4d)} & =-\frac{i}{st}\la12\ra^{2}[34]^{2}=i \frac{ \la12 \ra^{3}}{\la23\ra \la34\ra \la41\ra} \, .
 \end{flalign}
 Next, we consider the five-point amplitude   \eqref{5pttree}
  and reduce to a four-dimensional 
 helicity configuration $(1^{+},2^{+},3^{+},4^{-},5^{-})$. 
 For this case, only a few terms in  \eqref{5pttree} survive.
 The $\mathcal{A}$-tensor becomes
 \begin{align}
\mathcal{A}_{a\dot{a}b\dot{b}c\dot{c}d\dot{d}e\dot{e}}=\langle1_{a}|\hap_{2}\hap_{3}\hap_{4}\hap_{5}|1_{\dot{a}}]\langle2_{b}3_{c}4_{d}5_{e}\rangle[2_{\dot{b}}3_{\dot{c}}4_{\dot{d}}5_{\dot{e}}] \nonumber
 \\+\langle2_{b}|\hap_{3}\hap_{4}\hap_{5}\hap_{1}|2_{\dot{b}}]\langle3_{c}4_{d}5_{e}1_{a}\rangle[3_{\dot{c}}4_{\dot{d}}5_{\dot{e}}1_{\dot{a}}] \nonumber \\
+\langle3_{c}|\hap_{4}\hap_{5}\hap_{1}\hap_{2}|3_{\dot{c}}]\langle4_{d}5_{e}1_{a}2_{b}\rangle[4_{\dot{d}}5_{\dot{e}}1_{\dot{a}}2_{\dot{b}}]
\ , 
\end{align}
which in the four-dimensional limit  takes the form
 \beqa
 \label{aaa}
\mathcal{A}_{a\dot{a}b\dot{b}c\dot{c}d\dot{d}e\dot{e}}    &\overset{4d}{\longrightarrow} &  -[12]\langle23\rangle[34]\langle45\rangle[51]\times[23]^{2}\langle45\rangle^{2} \label{1staterm} \nonumber \\
 && -[23]\langle34\rangle[45]\langle51\rangle[12]\times[31]^{2}\langle45\rangle^{2} \label{2ndaterm} \nonumber \\
 && -[34]\langle45\rangle[51]\langle12\rangle[23]\times[12]^{2}\langle45\rangle^{2} \label{3rdaterm}\,.
\eeqa
For our specific helicity choice, the non-zero parts of the $\mathcal{D}$-tensor are those involving the Lorentz invariant brackets
\beq
\langle2_{b}3_{c}4_{d}5_{e}\rangle[1_{\dot{a}}3_{\dot{c}}4_{\dot{d}}5_{\dot{e}}]\ , 
\eeq
and
\beq
[2_{\dot{b}}3_{\dot{c}}4_{\dot{d}}5_{\dot{e}}]\langle1_{a}3_{c}4_{d}5_{e}\rangle\ , 
\eeq
where both reduce to  $[23]\langle45\rangle[13]\langle45\rangle$ in four dimensions. Each factor multiplies $\langle1_{a}(2.\tilde{\Delta}_{2})_{\dot{b}}]$ and $[1_{\dot{a}}(2.\Delta_{2})_{b}\rangle$. The quantities $\Delta_{i}$'s that are of interest here take the form:
\begin{eqnarray}
\Delta_{2}=\langle2|\hap_{3}\hap_{4}\hap_{5}-\hap_{5}\hap_{4}\hap_{3}|2\rangle\ ,  & \mathrm{and} & \tilde{\Delta}_{2}=[2|\hap_{3}\hap_{4}\hap_{5}-\hap_{5}\hap_{4}\hap_{3}|2]\,.
\end{eqnarray}
 Expanding the expression of the first non-vanishing $\mathcal{D}$-term
yields, 
\beqa
\langle1_{a}(2.\tilde{\Delta}_{2})_{\dot{b}}]&=&\langle1_{a}|2^{\dot{b}}][2_{\dot{b}}|3^{c}\rangle\langle3_{c}|4^{\dot{d}}][4_{\dot{d}}|5^{e}\rangle\langle5_{e}|2_{\dot{b}}]-\langle1_{a}|2^{\dot{b}}][2_{\dot{b}}|5^{e}\rangle\langle5_{e}|4^{\dot{d}}][4_{\dot{d}}|3^{c}\rangle\langle3_{c}|2_{\dot{b}}]
\nonumber \\
&\overset{4d}{\longrightarrow} & [12]\langle23\rangle[34]\langle45\rangle[52] - [12]\langle25\rangle[54]\langle43\rangle[32]
\ . 
\eeqa
Using similar  manipulations one can reduce  the $[1_{\dot{a}}(2.\Delta_{2})_{b}\rangle$ term. 
This yields:
\be
\label{eq:ddterm}
2\mathcal{D}_{a\dot{a}b\dot{b}c\dot{c}d\dot{d}e\dot{e}}\overset{4d}{\to}2\left([12]\langle23\rangle[34]\langle45\rangle[52]-[12]\langle25\rangle[54]\langle43\rangle[32]\right)\times[13][23]\langle45\rangle^{2}.
\ee
Combining \eqref{aaa} and \eqref{eq:ddterm}, one finds, after a little algebra, the expected Parke-Taylor result.

%%%%%%%%%%%%%%%%%%%%%%%%%%%%%%%%%%%%%

\newpage
\vspace{1cm}
%\newpage

\end{document}